\documentclass[12pt]{article}
\usepackage{amsmath,amssymb,epsfig}
\usepackage{axodraw}

\def\be{\begin{eqnarray}}
\def\ee{\end{eqnarray}}
\newcommand{\ep}{\varepsilon}
\begin{document}

\begin{center}
{\Large \bf Anomalous dimensions of twist-2 conformal operators \\
in supersymmetric Wess-Zumino model.}
\\ \vspace*{5mm} A.I.~Onishchenko$^{a,b}$
and V.N.~Velizhanin$^{c}$
\end{center}

\begin{center}
a) Department of Physics and Astronomy \\ Wayne State University,
Detroit, MI 48201, USA \\
\vspace*{0.5cm}
b) Institute for Theoretical and Experimental
Physics, \\ Moscow, Russia \\
\vspace*{0.5cm}
c) Theoretical Physics Department, \\ Petersburg Nuclear Physics Institute \\
Orlova Rosha, Gatchina, \\
188300, St. Petersburg, Russia
\end{center}

\begin{abstract}
In the present paper we are studying scale properties of twist-2 conformal operators
in supersymmetric Wess-Zumino model. In particular, we are interested in a construction
of multiplicatively renormalized conformal operators. We show, that in order to find
multiplicatively renormalized operators in this model, it is sufficient to find multiplicatively
renormalized operators only for one member of operator supermultiplet. We found a closed
analytical solution for fermionic conformal operator, which together with supersymmetry
transformations could be used to find remained multiplicatively renormalized bosonic
operators. Moreover, we found, that the knowledge of fermionic diagonal and non-diagonal anomalous
dimensions matrices allows us completely reconstruct the forward anomalous dimensions matrix in singlet case.
\end{abstract}

\section{Introduction}

Parton distributions in QCD satisfy the Balitsky-Fadin-Kuraev-Lipatov (BFKL)~\cite{BFKL}
and Dokshitzer-Gribov-Lipatov-Altarelli-Parizi~\cite{DGLAP} equations. When considered
in supersymmetric limit these equations reveal some remarkable properties. For example,
in the maximally supersymmetric  $N=4$ Yang-Mills theory there is a deep relation between BFKL and
DGLAP evolution equations~\cite{KL}. In particular, the anomalous dimensions of Wilson twist-2 operators
in $N=4$ SYM could be found from the eigenvalues of the kernel of BFKL equation. Moreover,
the anomalous dimensions matrix of the twist-2 operators in $N=4$ Supersymmetric Yang-Mills theory has
a very simple form with its elements being expressed in terms of one universal anomalous
dimension with shifted arguments~\cite{KL,KLV}. In leading order, universal anomalous dimension
was found to be proportional to the $\Psi$-function, what may be considered as an argument
in favor of integrability of this model~\cite{Integr}. Generally, the conformal invariance of the theory
in leading order allows us to construct multiplicatively  renormalized quasi-partonic operators~\cite{CPO} up to
this order in perturbation theory. However, in next-to-leading order in perturbation theory
multiplicative renormalization of conformal operators is violated due to necessity in regularization
of arising ultraviolet divergences.

The Efremov-Radyushkin-Brodsky-Lepage (ER-BL) equation~\cite{ERBL} may be viewed as some kind of the generalization of
the DGLAP equation for the case of non-forward distribution functions. In this case in and out hadronic
states in matrix elements of conformal operators are different, what allows us to study scale properties of
hadron wave functions. In the latter case the matrix elements of corresponding conformal operators are
considered between vacuum and hadronic state. It is known, that the eigenfunctions of ER-BL evolution equation are directly
related to multiplicatively renormalized conformal operators. Moreover, knowing ER-BL evolution kernels one may
find both diagonal and non-diagonal parts of anomalous dimensions matrix for corresponding conformal
operators.

Recently, there was a great interest in properties of before
mentioned operators in the context of famous AdS/CFT correspondence \cite{MGKPW}. Namely, there are some
calculations of the anomalous dimensions of such operators in the limit of large $j$ (Lorentz spin) from
both sides of the AdS/CFT correspondence \cite{GKP,FT,KLV,BGK}. There are, also some predictions for anomalous
dimensions of other types of operators from string theory \cite{FT03}, partially confirmed by field theory
calculations. It should be noted, that up to this moment only diagonal part of anomalous dimension matrix have
been studied in the context of AdS/CFT correspondence and it would be interesting to compare the results for
its non-diagonal part with the appropriate result from string theory. So, we believe, that non-diagonal parts
of anomalous dimensions matrices in supersymmetric theories may reveal some interesting properties and thus
are worth to study.

In this paper we start our study with a simple supersymmetric model - supersymmetric Wess-Zumino
model~\cite{WZmodel}
which contains only matter superfields. This model allow us to understand  general properties of the
ER-BL equations and their solutions in the supersymmetric theories. As the next step we suppose
to perform a similar analysis for field theories with more supersymmetries. We will be mainly interested
in the construction of multiplicatively renormalized conformal operators for non-forward kinematics and
subsequent understanding of their string counterparts. This will be a subject of one of our next papers.

To be precise, in the present work we calculate evolution kernels for non-singlet and singlet unpolarized
conformal operators together with similar quantities in polarized case. We study the transformation
properties of twist-2 conformal operators in Wess-Zumino model under supersymmetry transformations. At this point,
it is natural to introduce fermionic operators (with respect to quantum numbers), as components of corresponding
supersymmetric chiral multiplets. It is shown, that the latter allow us reduce the problem of finding the solution of
evolution equations in singlet case (here we have operator mixing adding extra complexity) to the similar
problem for fermionic operators. The corresponding problem for fermionic operators turns out to be much
simpler and we will be able to obtain a closed analytical solution for multiplicatively renormalized fermionic operators,
which together with supersymmetry transformation could used to obtain a solution of evolution equations in
the singlet case. Moreover, we found, that the knowledge of fermionic diagonal and non-diagonal anomalous
dimensions matrices allows us completely reconstruct the forward anomalous dimensions matrix in singlet case.

Our present consideration is based on two approaches developed for studying non-forward partonic
distribution functions. The first one follows the articles of Mikhailov and Radyushkin
\cite{MRPhi36,MRQCD,MRKernel}, where an effective method for calculation of the
evolution kernels of ER-BL equation was proposed \cite{MRPhi36,MRQCD}. The properties of evolution
kernels were later studied in \cite{MRKernel}. The absence of the gauge fields in our problem
considerably simplifies the whole calculation, so here, we are following Ref.~\cite{MRPhi36}, where
the scalar $\phi^3_{(6)}$ theory in the space-time dimension six was considered. The second
approach is based on papers of M\"uller~\cite{MPhi3,MQEDNS},
where the renormalized conformal Ward identities were extensively exploited
to find non-diagonal parts of anomalous dimension.

The present paper is organized as follows. In section 2, after briefly recalling some necessary
background about distribution amplitudes, conformal operators and supersymmetric Wess-Zumino
model, we give our results for ER-BL evolution kernels in unpolarized and polarized cases
together with some comments on the details
of this calculation. In section 3 we introduce an approach for the evaluation of non-diagonal parts
of anomalous dimensions matrices for conformal operators, we are interested in, based on the analysis of conformal
Ward identities for Green functions with operator insertions. The obtained non-diagonal anomalous
dimensions matrices were checked to coincide with those coming from ER-BL evolution kernels evaluated before.
In section 4, we combine our conformal operators into one supersymmetric
chiral multiplet and study the constrains on anomalous dimensions of conformal operators following
from supersymmetric Ward identity. Next, in section 5 we construct multiplicatively renormalized
fermionic operator up to next-to-leading order in perturbation theory and explain how corresponding
solutions could be found for other operators. Finally, section 6 contains our conclusion.

\section{ER-BL evolution kernels in the supersymmetric Wess-Zumino model}

\subsection{Preliminary}

The field-theoretical background for the study of the distribution amplitudes is
provided by their relation to the matrix elements of non-local operators sandwiched
between hadron and vacuum states
\begin{equation}
\phi(x)=\frac{1}{2\pi}\int dz_-e^{i x p_+z_-}<0|\varphi(0)\varphi(z_-)|h(p_+)>,
\end{equation}
where $p_+$ is hadron momentum, $x$ - momentum fraction carried by field $\varphi$ and $z_-$ stands
for light-cone coordinate.

ER-BL equation describes
the scale dependence of the distribution amplitudes (DA) or more general non-forward parton distribution
in contrast to scale dependence of forward distribution functions governed by DGLAP equation. The ER-BL
equation is generally written as
\begin{equation}\label{EvEq}
\mu^2\frac{d}{d\mu^2}\phi(x,\mu)=\int^1_0 d y {\mbox {\bf V}}(x,y|\alpha(\mu))\phi(y,\mu)\, ,
\end{equation}
where evolution kernel
\begin{equation}\label{EvKer}
{\mbox {\bf V}}(x,y|\alpha)=\frac{\alpha}{4\pi}{\mbox {\bf V}}^{(0)}(x,y)
+\left(\frac{\alpha}{4\pi}\right)^2{\mbox {\bf V}}^{(1)}(x,y)+...
\end{equation}
is given as a series in the coupling constant and is calculated in perturbation theory.
In leading order the solution of evolution equation (\ref{EvEq}) could be constructed
introducing moments of DA with respect to Gegenbauer polynomials in the case of
bosonic (on quantum numbers) DA or Jacobi polynomials in the case of fermionic DA.
For example, for non-singlet bosonic DA we have:
\be
\int_0^1 dx C_k^{3/2}(x-\bar x)\phi (x, Q^2) =
\langle 0|{\cal O}_{k, k}^{\psi} (\mu^2)|P\rangle_{\mu^2 = Q^2}^{\mathbf red},
\ee
where $\bar x=1-x$ and
$\langle 0|{\cal O}_{k, k}^{\psi} (\mu^2)|P\rangle^{\mathbf red} =
\langle 0|{\cal O}_{k, k}^{\psi} (\mu^2)|P\rangle /P_{+}^{k+1}$ denotes reduced expectation
value of non-singlet bosonic conformal twist-2 operator:
\be
{\mathcal O}^{\psi}_{j,l}=
\frac12\bar \psi_+ (i\partial_+)^l\gamma_+
C^{3/2}_j\left(\frac{{\mathcal D}_+}{\partial_+}\right)\psi_+
\ee
where
${\mathcal D}=\overrightarrow{\partial}- \overleftarrow{\partial}\!$,
$\partial=\overrightarrow{\partial}+\overleftarrow{\partial}\!$.
In leading order this and other conformal operators, we will consider later, are multiplicatively
renormalized, so that the solution of evolution equation can be easily found. The evolution
equation for DA (\ref{EvEq}) leads to renormalization group equations for corresponding
conformal operators:
\be
\mu\frac{d}{d\mu}{\cal O}_{j, l}^{\psi} (\mu^2) =
\sum_{k=0}^{l}\gamma_{j, k} (\alpha (\mu^2)) {\cal O}_{k, l}^{\psi }(\mu^2),
\ee
where the anomalous dimensions matrix
\be
\gamma_{j, k} =
\left(\frac{\alpha}{4\pi}\right)\gamma_k^{(0)}\delta_{j, k} +
\left(\frac{\alpha}{4\pi}\right)^2\gamma_{j, k}^{(1)} + \ldots
\ee
is diagonal in leading order and could be extracted from the expression for evolution
kernel:
\begin{equation}
\int^1_0 d xC^{3/2}_j(x-\bar x){\mbox {\bf V}}(x,y|\alpha)=
\sum_{k=0}^j\gamma_{jk}(\alpha)C^{3/2}_k(y-\bar y).\label{EKADRel}
\end{equation}

So, we see that the problem of determining the evolution kernels for DA and the
problem of calculation of anomalous dimensions matrix for corresponding conformal
operators are closely related. In what follows, we are using two different
approaches to determine DA evolution kernels and anomalous dimensions matrices of
our conformal operators. Equations, similar to (\ref{EKADRel}) will be used only to check,
that both methods give us the same results.

\subsection{Wess-Zumino model}
The field content of supersymmetric Wess-Zumino model~\cite{WZmodel} consists of the (anti)chiral supermultiplets
containing off mass shell a complex scalar field $\varphi$, one Majorana fermion field $\psi$
(``quark'' field) and a complex auxiliary scalar field $F$:
\begin{eqnarray}
\Phi(y,\theta) & = & \varphi(y) + \sqrt{2}\theta\psi(y) + \theta\theta {\mathcal F}(y) \label{ChSM}\\
\Phi(x,\theta,\bar\theta) & = & \varphi(x) + \sqrt{2} \theta\psi(x)
+ \theta\theta {\mathcal F}(x)\nonumber  \\
& - &  i \theta\sigma^\mu\bar\theta \partial_\mu \varphi(x) - \frac14
\theta\theta\bar\theta\bar\theta\partial_\mu\partial^\mu \varphi(x)
- \frac{i}{\sqrt{2}} \theta\theta\bar\theta\bar\sigma_\mu
\partial^\mu\psi(x)
\end{eqnarray}
The lagrangian for this model is conveniently written using superfield formalism
and is given by
\begin{equation}
{\cal L} = \int d^2\theta d^2\bar\theta\bar\Phi_i \Phi^i
-\frac{1}{3!}\int d^2\theta Y_{ijk} \Phi^i \Phi^j \Phi^k
-\frac{1}{3!}\int d^2\bar\theta \bar Y^{ijk} \bar\Phi_i \bar\Phi_j \bar\Phi_k\,,
\label{Lagrangian}
\end{equation}
where $Y_{ijk}$ is Yukawa coupling, symmetric with respect to all indices. The
first term in the lagrangian stands for kinetic terms of all fields present
in this model, second and third terms denote superpotential
(and its hermitian conjugate), which remains non-renormalized to all orders in
perturbative theory. Due to non-renormalization of the superpotential the $\beta$-function
for Yukawa coupling completely determined by the wave function renormalization of fields.
To simplify notation it is convenient to put $Y^{ijk}=g$. The first coefficient of
Yukawa $\beta$-function $\beta(g^2)= b_0 \frac{g^3}{(4\pi)^2} + \ldots$ in Wess-Zumino model is $b_0=-3$
and the leading order anomalous dimension of the chiral superfield is $\gamma_\Phi=\frac{g^2}{(4\pi)^2}$.

In what follows, we will also need this lagrangian written in
component notation. Rewriting complex scalar field $\varphi=(A+i B)/\sqrt{2}$ in terms of
real scalar $A$ and pseudoscalar $B$ fields, complex auxiliary
scalar field ${\cal F}=(F+i G)/\sqrt{2}$ in terms of real scalar $F$ and pseudoscalar $G$ auxiliary
fields the Wess-Zumino lagrangian takes the form
\begin{eqnarray}
{\cal L} &=&
\frac12\partial^\mu A\partial_\mu A
+\frac12\partial^\mu B\partial_\mu B
+\frac{i}{2}\bar\psi\gamma^\mu\partial_\mu\psi
+\frac12\left(F^2+G^2\right)\nonumber\\
&+&g\left[F (A^2+B^2)+2GAB
-i\bar\psi(A+\gamma_5 B)\psi\right].
\label{CompLag}
\end{eqnarray}
The supersymmetric transformations for the component fields could be summarized
as follows
\begin{eqnarray}
\delta^Q A&=&\bar\epsilon\psi\;,\qquad\qquad
\delta^Q B=\bar\epsilon\gamma_5\psi\;,\nonumber\\
\delta^Q\psi&=&i\gamma^\mu\partial_\mu(A+\gamma_5 B)\epsilon
+i\left(F+\gamma_5 G\right)\epsilon\;,\\
\delta^Q F&=&\bar\epsilon\gamma^\mu\partial_\mu\psi\;,\qquad
\delta^Q G=\bar\epsilon\gamma^\mu\gamma_5\partial_\mu\psi\;.\nonumber
\label{SUSYTransf}
\end{eqnarray}
After eliminating auxiliary fields $F$ and $G$ with the use of their equations of motion
the lagrangian in Eq.~(\ref{CompLag}) becomes
\begin{eqnarray}
{\cal L} &=&
\frac12\partial^\mu A\partial_\mu A
+\frac12\partial^\mu B\partial_\mu B
+\frac{i}{2}\bar\psi\gamma^\mu\partial_\mu\psi\nonumber\\
&-&\frac12 g^2\left(A^2+B^2\right)^2
-ig\bar\psi\left( A+\gamma_5 B\right)\psi.
\label{CompLagEl}
\end{eqnarray}
Without auxiliary fields supersymmetric transformations (\ref{SUSYTransf})
do not form a closed algebra. However, introducing light-cone coordinates
and defining light-cone fields one can find closed superalgebra for
``plus'' - components of the fields. In light-cone coordinates general four-vector
$X^\mu$ is represented as
$X^{\pm}=(1/\sqrt{2})(X^0\pm X^3)$, and $X^\mu Y_\mu =X_+Y_-+X_-Y_+-X^i Y^i$, where
i=1,2. For spinors the appropriate definition is $\lambda_\pm=\frac12 \gamma_\pm\gamma_\mp\lambda$,
so that $\lambda=\lambda_++\lambda_-$. Then choosing $\epsilon=\epsilon_-$ the supersymmetric transformations take the following form~\cite{LightConeSUSY}
\begin{eqnarray}
\delta^Q A&=&\bar\epsilon_-\psi_+\;,\nonumber\\
\delta^Q B&=&\bar\epsilon_-\gamma_5\psi_+\;,\\
\delta^Q\psi_+&=&i\partial_+(A-\gamma_5 B)\gamma_-\epsilon_-\;.\nonumber
\label{SUSYTransfLightCone}
\end{eqnarray}

Now, let us  introduce the local conformal Wilson twist-2 operators appearing in this model
for unpolarized and polarized cases
\begin{eqnarray}
{\mathcal O}^{\psi}_{j,l}&=&
\frac12\bar \psi_+ (i\partial_+)^l\gamma_+
C^{3/2}_j\left(\frac{{\mathcal D}_+}{\partial_+}\right)\psi_+\,,\label{qqn}\\
{\widetilde {\mathcal O}}^{\psi}_{j,l}&=&
\frac12\bar \psi_+ (i\partial_+)^l\gamma_+\gamma_5
C^{3/2}_j\left(\frac{{\mathcal D}_+}{\partial_+}\right)\psi_+\,,\label{qqp}\\
{\mathcal O}^{\phi}_{j,l}&=&
A(i\partial_+)^{l+1}C^{1/2}_{j+1}\left(\frac{{\mathcal D}_+}{\partial_+}\right)A
+B(i\partial_+)^{l+1}C^{1/2}_{j+1}\left(\frac{{\mathcal D}_+}{\partial_+}\right)B\,,\label{ssn}\\
{\widetilde {\mathcal O}}^{\phi}_{j,l}&=&
A(i\partial_+)^{l+1}C^{1/2}_{j+1}\left(\frac{{\mathcal D}_+}{\partial_+}\right)B
+B(i\partial_+)^{l+1}C^{1/2}_{j+1}\left(\frac{{\mathcal D}_+}{\partial_+}\right)A
\,,\label{ssp}
\end{eqnarray}
where
${\mathcal D}=\overrightarrow{\partial}- \overleftarrow{\partial}\!$,
$\partial=\overrightarrow{\partial}+\overleftarrow{\partial}\!$ and
$C^{\nu}_n(z)$ - Gegenbauer polynomials
\begin{equation}
C^{\nu}_n(z)=\frac{(-1)^n 2^n}{n!}\frac{\Gamma(n+\nu)}{\Gamma(\nu)}
\frac{\Gamma(n+2\nu)}{\Gamma(2n+2\nu)}(1-z^2)^{-\nu+1/2}
\frac{d^n}{dz^n}\left[(1-z^2)^{n+\nu-1/2}\right].
\end{equation}
Moreover in this model one can introduce also the following fermionic (by quantum
numbers) operators
\begin{eqnarray}
{\mathcal O}^{\mathbf {fer}}_{j,l}&=&
\bar\psi_+(i\partial_+)^{l+1}
P^{(1,0)}_{j+1}\left(\frac{{\mathcal D}_+}{\partial_+}\right)(A+\gamma_5B)\label{ConfOpFWZ1}\,,\\
{\widetilde{\mathcal O}}^{\mathbf {fer}}_{j,l}&=&
\bar\psi_+(i\partial_+)^{l+1}\gamma_5
P^{(1,0)}_{j+1}\left(\frac{{\mathcal D}_+}{\partial_+}\right)(A+\gamma_5B)\,,\label{ConfOpFWZ2}
\end{eqnarray}
where
$P^{(\alpha,\beta)}_n(z)$ - Jacobi polynomials
\begin{equation}
P^{(\alpha,\beta)}_n(z)=\frac{(-1)^n}{n!\;2^n}(1-z)^{-\alpha}
(1+z)^{-\beta}
\frac{d^n}{dz^n}\left[(1-z)^\alpha(1+z)^\beta(1-z^2)^n\right].
\end{equation}

As was already mentioned in introduction, we will be interested in the renormalization
properties of these operators. It should be noted, that in the singlet case there is mixing
between bosonic operators formed by fermion and scalar fields. Also, in the non-forward kinematics,
in contrast to forward case, the operators (\ref{qqn}) and (\ref{ssn}) will mix under renormalization
not only with each other, but also with the total derivatives of these operators.

The scale properties of twist-2 conformal operators, we are interested in, could be deduced
from scale properties of corresponding distribution amplitudes. So, in what follows, we will
first determine the evolution kernels for DA in the cases of interest.

\subsection{Unpolarized case}

\subsubsection{Non-singlet case}

\begin{figure}[ht]
\vspace{-1.5cm}
\begin{center}
\begin{picture}(680,120)(0,0)
\SetScale{0.65}
\SetWidth{1}
\Line(40,-20)(105,109)
\Line(160,-20)(95,109)
\CArc(100,100)(10,0,360)
\DashLine(122,55)(78,55){7}
\DashLine(142,15)(58,15){7}
\Vertex(122,55){3}
\Vertex(78,55){3}
\Vertex(142,15){3}
\Vertex(58,15){3}
\Text(65,-25)[c]{(a)}

\Line(240,-20)(305,109)
\Line(360,-20)(295,109)
\CArc(300,100)(10,0,360)
\DashCArc(278,55)(20,248,61){7}
\DashLine(342,15)(258,15){7}
\Vertex(286,72){3}
\Vertex(269,38){3}
\Vertex(342,15){3}
\Vertex(258,15){3}
\Text(195,-25)[c]{(b)}

\Line(440,-20)(505,109)
\Line(560,-20)(495,109)
\CArc(500,100)(10,0,360)
\CArc(500,15)(20,0,360)
\DashLine(542,15)(520,15){7}
\DashLine(480,15)(458,15){7}
\Vertex(520,15){3}
\Vertex(480,15){3}
\Vertex(542,15){3}
\Vertex(458,15){3}
\Text(325,-25)[c]{(c)}

\Line(678,55)(705,109)
\Line(640,-20)(658,15)
\Line(722,55)(695,109)
\Line(760,-20)(742,15)
\CArc(700,100)(10,0,360)
\Line(722,55)(678,55)
\DashLine(722,55)(742,15){7}
\Line(742,15)(658,15)
\DashLine(678,55)(658,15){7}
\Vertex(722,55){3}
\Vertex(678,55){3}
\Vertex(742,15){3}
\Vertex(658,15){3}
\Text(455,-25)[c]{(d)}
\end{picture}
\end{center}
\vspace*{5mm}
\caption{Two-loop diagrams for quark operator Eq.~(\ref{qqn}) sandwiched
between quark states}
\label{TwoLoopQQ}
\end{figure}
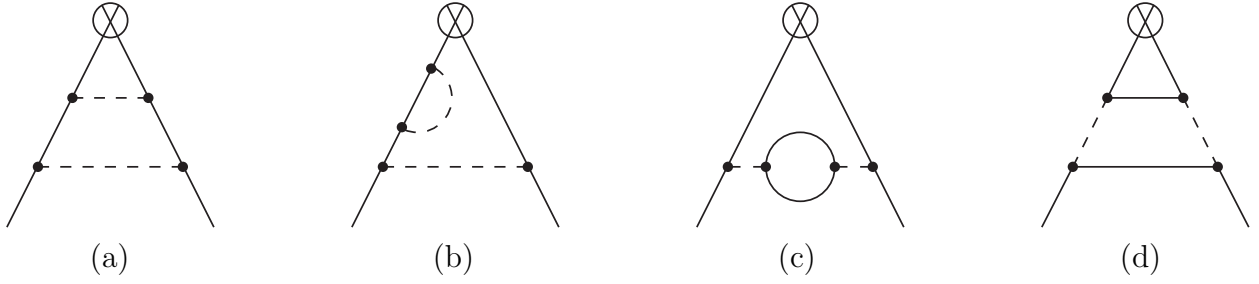

In non-singlet case we need to calculate the evolution kernel of DA, which Gegenbauer moments
are given by matrix elements of operator~(\ref{qqn}) sandwiched between quark states.
In leading order there is only one diagram shown in Fig.~\ref{OneLoop}.a and in next-to-leading order
there are three diagrams shown
in Fig.~\ref{TwoLoopQQ}.a,~\ref{TwoLoopQQ}.b and ~\ref{TwoLoopQQ}.c~\footnote{One
need to include also graphs with external self-energies which give $\delta$-function
contribution in the coordinate space}. For the diagram evaluation we used the method proposed by Mikhailov
and Radyushkin \cite{MRPhi36,MRQCD}. The calculation details together with diagram by diagram answers
may be found in Appendices A and B. The final result for the non-singlet non-forward evolution kernel is
($\alpha=\frac{g^2}{4\pi}$)
\begin{eqnarray}
{\mbox {\bf V}}_{\!\! NS}(x,y)&=&\frac{\alpha}{4 \pi}{\mbox {\bf V}}^{(0)}_{\!\! NS}(x,y)+
\left(\frac{\alpha}{4 \pi}\right)^2{\mbox {\bf V}}^{(1)}_{\!\! NS}(x,y)\\
{\mbox {\bf V}}^{(0)}_{\!\! NS}(x,y)&=& \theta(x<y)\left\{2\frac{x}{y}-\delta(x-y)\right\}+
  \left(\begin{array}{c}
    x \leftrightarrow \bar{x}\\
    y \leftrightarrow \bar{y}
  \end{array}\right)\\
{\mbox {\bf V}}^{(1)}_{\!\! NS}(x,y)&=& \theta(x<y)\left\{
-4 \frac{x}{y}
- \frac{\bar{x}}{\bar{y}} \ln ( \bar{x} )
- \frac{x}{y} \ln ( x )
+ \frac{\bar{x}}{\bar{y}} \ln \left( 1 - \frac{x}{y} \right)
- \frac{x}{y} \ln \left( 1 - \frac{x}{y} \right)\right. \nonumber\\
&+& \left.\frac{\bar{x}}{y} \ln^2 ( \bar x )
+ 2 \frac{x}{\bar{y}} \ln ( x ) \ln ( y )
- \frac{x}{\bar{y}} \ln^2 ( y )
+\delta(x-y)\right\}
+\left(\begin{array}{c}
    x \leftrightarrow \bar{x}\\
    y \leftrightarrow \bar{y}\label{KernelNS}
  \end{array}\right)
\end{eqnarray}
In the forward case the evolution kernels are given by
\begin{eqnarray}
{\mbox {\bf P}}^{(0)}_{NS}(z)&=&2\bar z-\delta(1-z)\\
{\mbox {\bf P}}^{(1)}_{NS}(z)&=&-4\bar z-2\bar z\ln(\bar z)+2\bar z\ln (z)+(1+z)\ln^2(z)+\delta(1-z)
\end{eqnarray}
First three terms in the second line in Eq.~(\ref{KernelNS}) represent non-symmetric piece of
the kernel $y\bar y{\mbox {\bf V}}_{\!\! NS}(x,y)$ with respect to the exchange of variables
($  x \leftrightarrow y $)
\begin{equation}
y\bar y{\mbox {\bf V}}_{\!\! NS}(x,y) = x\bar x{\mbox {\bf V}}_{\!\! NS}(y,x)
\end{equation}
and, as was shown in Ref.~\cite{MRPhi36,MRKernel}, lead to non-diagonal part of anomalous dimensions matrix in the
basis of Gegenbauer polynomials (see below).

\subsubsection{Singlet case}

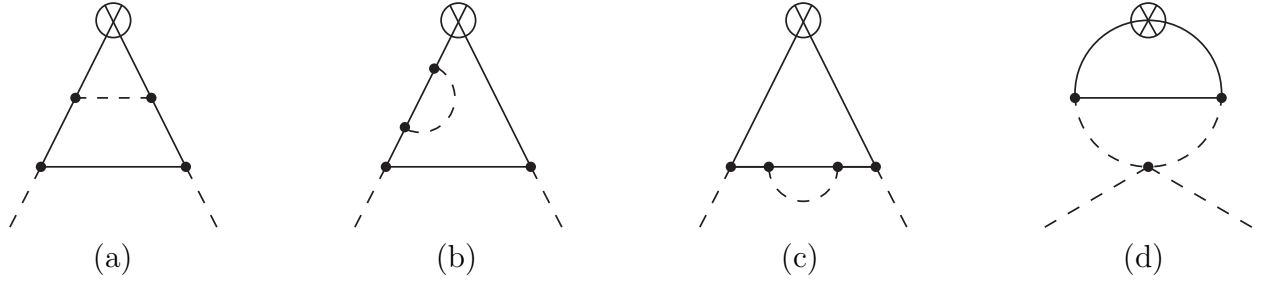
\begin{figure}[ht]
\vspace{-1.5cm}
\begin{center}
\begin{picture}(680,120)(0,0)
\SetScale{0.65}
\SetWidth{1}
\DashLine(40,-20)(58,15){7}
\Line(78,55)(58,15)
\Line(78,55)(105,109)
\DashLine(160,-20)(142,15){7}
\Line(122,55)(142,15)
\Line(122,55)(95,109)
\CArc(100,100)(10,0,360)
\DashLine(122,55)(78,55){7}
\Line(142,15)(58,15)
\Vertex(122,55){3}
\Vertex(78,55){3}
\Vertex(142,15){3}
\Vertex(58,15){3}
\Text(65,-25)[c]{(a)}

\Line(258,15)(305,109)
\DashLine(240,-20)(258,15){7}
\Line(342,15)(295,109)
\DashLine(360,-20)(342,15){7}
\CArc(300,100)(10,0,360)
\DashCArc(278,55)(20,248,61){7}
\Line(342,15)(258,15)
\Vertex(286,72){3}
\Vertex(269,38){3}
\Vertex(342,15){3}
\Vertex(258,15){3}
\Text(195,-25)[c]{(b)}

\DashLine(440,-20)(458,15){7}
\Line(458,15)(505,109)
\DashLine(560,-20)(542,15){7}
\Line(542,15)(495,109)
\CArc(500,100)(10,0,360)
\DashCArc(500,15)(20,180,360){7}
\Line(542,15)(520,15)
\Line(542,15)(458,15)
\Line(480,15)(458,15)
\Vertex(520,15){3}
\Vertex(480,15){3}
\Vertex(542,15){3}
\Vertex(458,15){3}
\Text(325,-25)[c]{(c)}

\Line(695,91)(705,109)
\DashLine(640,-20)(700,15){7}
\Line(705,91)(695,109)
\DashLine(760,-20)(700,15){7}
\DashCArc(700,57.5)(42.5,180,360){7}
\CArc(700,57.5)(42.5,0,180)
\CArc(700,100)(10,0,360)
\Line(742.5,55)(657.5,55)
\Vertex(742.5,55){3}
\Vertex(657.5,55){3}
\Vertex(700,15){3}
\Text(455,-25)[c]{(d)}
\end{picture}
\end{center}
\vspace*{5mm}
\caption{Two-loop diagrams for quark operator Eq.~(\ref{qqn}) sandwiched
between scalar states}
\label{TwoLoopQS}
\end{figure}

\begin{figure}[ht]
\vspace{-1.5cm}
\begin{center}
\begin{picture}(680,120)(0,0)
\SetScale{0.65}
\SetWidth{1}
\Line(40,-20)(58,15)
\Line(78,55)(58,15)
\Line(96,91)(105,109)
\Line(95,109)(104,91)
\DashLine(78,55)(105,109){7}
\Line(160,-20)(142,15)
\Line(122,55)(142,15)
\DashLine(122,55)(95,109){7}
\CArc(100,100)(10,0,360)
\Line(122,55)(78,55)
\DashLine(142,15)(58,15){7}
\Vertex(122,55){3}
\Vertex(78,55){3}
\Vertex(142,15){3}
\Vertex(58,15){3}
\Text(65,-25)[c]{(a)}

\DashLine(305,109)(286,72){7}
\Line(305,109)(296,91)
\Line(304,91)(295,109)
\DashLine(258,15)(269,38){7}
\Line(240,-20)(258,15)
\DashLine(342,15)(295,109){7}
\Line(360,-20)(342,15)
\CArc(300,100)(10,0,360)
\CArc(278,55)(20,0,360)
\Line(342,15)(258,15)
\Vertex(286,72){3}
\Vertex(269,38){3}
\Vertex(342,15){3}
\Vertex(258,15){3}
\Text(195,-25)[c]{(b)}

\Line(440,-20)(458,15)
\DashLine(458,15)(505,109){7}
\Line(496,91)(505,109)
\Line(560,-20)(542,15)
\DashLine(542,15)(495,109){7}
\Line(504,91)(495,109)
\CArc(500,100)(10,0,360)
\DashCArc(500,15)(20,180,360){7}
\Line(542,15)(520,15)
\Line(542,15)(458,15)
\Line(480,15)(458,15)
\Vertex(520,15){3}
\Vertex(480,15){3}
\Vertex(542,15){3}
\Vertex(458,15){3}
\Text(325,-25)[c]{(c)}

\Line(695,91)(705,109)
\Line(640,-20)(668,15)
\Line(705,91)(695,109)
\Line(760,-20)(732,15)
\DashCArc(700,77.5)(22.5,0,360){7}
\CArc(700,100)(10,0,360)
\DashLine(700,55)(732,15){7}
\Line(732,15)(668,15)
\DashLine(700,55)(668,15){7}
\Vertex(700,55){3}
\Vertex(732,15){3}
\Vertex(668,15){3}
\Text(455,-25)[c]{(d)}
\end{picture}
\end{center}
\vspace*{5mm}
\caption{Two-loop diagrams for scalar operator Eq.~(\ref{ssn}) sandwiched
between quark states}
\label{TwoLoopSQ}
\end{figure}
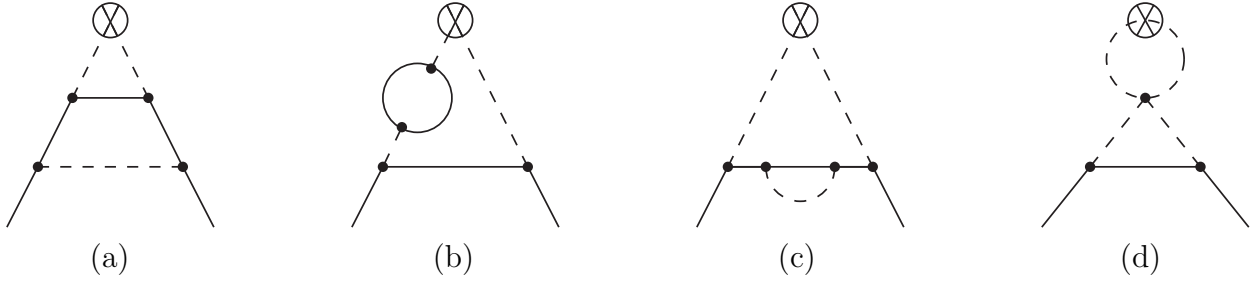

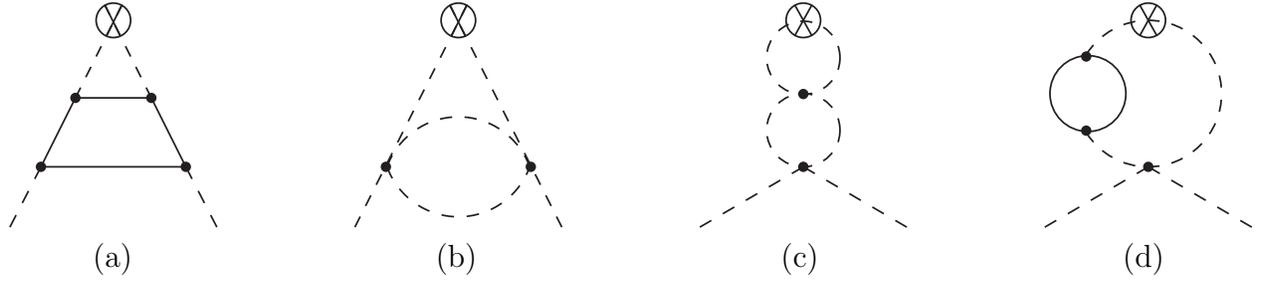
\begin{figure}[ht]
\vspace{-1.5cm}
\begin{center}
\begin{picture}(680,120)(0,0)
\SetScale{0.65}
\SetWidth{1}
\DashLine(40,-20)(58,15){7}
\Line(96,91)(105,109)
\Line(95,109)(104,91)
\Line(78,55)(58,15)
\DashLine(78,55)(105,109){7}
\DashLine(160,-20)(142,15){7}
\Line(122,55)(142,15)
\DashLine(122,55)(95,109){7}
\CArc(100,100)(10,0,360)
\Line(122,55)(78,55)
\Line(142,15)(58,15)
\Vertex(122,55){3}
\Vertex(78,55){3}
\Vertex(142,15){3}
\Vertex(58,15){3}
\Text(65,-25)[c]{(a)}

\Line(305,109)(296,91)
\DashLine(240,-20)(305,109){7}
\Line(304,91)(295,109)
\DashLine(360,-20)(295,109){7}
\CArc(300,100)(10,0,360)
\DashCArc(300,30)(44,200,-17){7}
\DashCArc(300,0)(44,20,160){7}
\Vertex(342,15){3}
\Vertex(258,15){3}
\Text(195,-25)[c]{(b)}

\Line(505,109)(495,91)
\DashLine(440,-20)(500,15){7}
\Line(505,91)(495,109)
\DashLine(560,-20)(500,15){7}
\CArc(500,100)(10,0,360)
\DashCArc(500,36.25)(21.25,0,360){7}
\DashCArc(500,78.25)(21.25,0,360){7}
\Vertex(500,15){3}
\Vertex(500,57.25){3}
\Text(325,-25)[c]{(c)}

\Line(695,91)(705,109)
\DashLine(640,-20)(700,15){7}
\Line(705,91)(695,109)
\DashLine(760,-20)(700,15){7}
\DashCArc(700,57.5)(42.5,212,148){7}
\CArc(665,57.5)(22,0,360)
\CArc(700,100)(10,0,360)
\Vertex(700,15){3}
\Vertex(664,79){3}
\Vertex(664,36){3}
\Text(455,-25)[c]{(d)}
\end{picture}
\end{center}
\vspace*{5mm}
\caption{Two-loop diagrams for scalar operator Eq.~(\ref{ssn}) sandwiched
between scalar states}
\label{TwoLoopSS}
\end{figure}

In singlet case there is mixing between operators (\ref{qqn}) and (\ref{ssn})
and the evolution kernel takes now the matrix form
\begin{eqnarray}
 {\mbox {\bf V}}(x,y)&=&
 \left(\begin{array}{cc}
  {\mbox{\bf V}}_{\!\!\psi \psi}(x,y)&{\mbox{\bf V}}_{\!\!\psi \varphi}(x,y)  \\[3mm]
  {\mbox{\bf V}}_{\!\!\varphi \psi}(x,y)&{\mbox{\bf V}}_{\!\!\varphi \varphi}(x,y)
  \end{array}\right)
\end{eqnarray}
with the elements of this matrix in leading order (diagrams Fig.~\ref{OneLoop}.a~-~\ref{OneLoop}.d)
given by
\begin{eqnarray}
  {\mbox{\bf V}}^{(0)}_{\!\!\psi \psi}(x,y) & = &\theta(x<y)\left\{
  2\frac{x}{y} -\delta (x-y)\right\}
  +  \left(\begin{array}{c}
    x \leftrightarrow \bar{x}\\
    y \leftrightarrow \bar{y}
  \end{array}\right)\nonumber\\
  {\mbox {\bf V}}^{(0)}_{\!\!\psi \varphi}(x,y) & = &\theta(x<y)
  \Bigl\{ 2x \Bigr\}
    -  \left(\begin{array}{c}
    x \leftrightarrow \bar{x}\\
    y \leftrightarrow \bar{y}
  \end{array}\right)\nonumber\\
  {\mbox {\bf V}}^{(0)}_{\!\!\varphi \psi}(x,y) & = &\theta(x<y)\left\{
  - \frac{2}{y}\right\}
    -  \left(\begin{array}{c}
    x \leftrightarrow \bar{x}\\
    y \leftrightarrow \bar{y}
  \end{array}\right)\nonumber\\
  {\mbox {\bf V}}^{(0)}_{\!\!\varphi \varphi}(x,y) & = &\theta(x<y)
  \Bigl\{- 2 -\delta (x-y)\Bigr\}
   +  \left(\begin{array}{c}
    x \leftrightarrow \bar{x}\\
    y \leftrightarrow \bar{y}
  \end{array}\right)
\end{eqnarray}
and next-to-leading orders (diagrams Figs.~\ref{TwoLoopQQ}-~\ref{TwoLoopSS})
\begin{eqnarray}
  {\mbox{\bf V}}^{(1)}_{\!\!\psi \psi}(x,y) & = &\theta(x<y)\left\{
  - 6 \frac{x}{y} + \frac{x}{y} \ln \left( \frac{x}{y} \right)
  - \frac{\bar{x}}{\bar{y}} \ln ( \bar{x} )
  - \frac{x}{y} \ln ( x )
  + \frac{1}{\bar y}\left(1-\frac{x}{y}\right) \ln \left( 1 - \frac{x}{y} \right) \right.\nonumber\\
  &+&\left.
  \frac{\bar{x}}{y} \ln^2 ( \bar{x} )
  + 2 \frac{x}{\bar{y}} \ln ( x ) \ln ( y )
  - \frac{x}{\bar{y}} \ln^2 ( y )
  +\delta (x-y)\right\}
  +  \left(\begin{array}{c}
    x \leftrightarrow \bar{x}\\
    y \leftrightarrow \bar{y}
  \end{array}\right)\label{EKQQNLO}\\
  {\mbox {\bf V}}^{(1)}_{\!\!\psi \varphi}(x,y) & = &\theta(x<y)\left\{
 -  6 x
 - x \ln \left( \frac{x}{y} \right)
 + 3 \bar{x}\ln ( \bar{x} )
 - 3 x \ln ( x )
 - \ln \left( 1 - \frac{x}{y} \right)\right. \nonumber\\
 &+& \left.\bar{x} \ln^2 ( \bar{x} )
 - 2 x \ln ( x ) \ln ( y )
 + x \ln^2 ( y )\right.\biggr\}
    -  \left(\begin{array}{c}
    x \leftrightarrow \bar{x}\\
    y \leftrightarrow \bar{y}
  \end{array}\right)\label{EKQSNLO}\\
  {\mbox {\bf V}}^{(1)}_{\!\!\varphi \psi}(x,y) & = &\theta(x<y)\left\{
  \frac{6}{y}
  + \frac{1}{y} \ln \left( \frac{x}{y}\right)
  - \frac{1}{\bar{y}} \ln ( \bar{x} )
  + \frac{1}{y} \ln ( x )
  + \frac{1}{y\bar{y}} \ln \left( 1 - \frac{x}{y} \right) \right.\nonumber\\
  &+&\left. 
   \frac{1}{y} \ln^2 ( \bar{x} )
  - \frac{2}{\bar{y}} \ln ( x ) \ln ( y )
  + \frac{1}{\bar{y}} \ln^2 ( y )\right\}
    -  \left(\begin{array}{c}
    x \leftrightarrow \bar{x}\\
    y \leftrightarrow \bar{y}
  \end{array}\right)\label{EKSQNLO}\\
  {\mbox {\bf V}}^{(1)}_{\!\!\varphi \varphi}(x,y) & = &\theta(x<y)\left\{
  2
  + \ln \left( \frac{x}{y} \right)
  + \ln^2 (\bar{x} )
  + 2 \ln ( x ) \ln ( y )
  - \ln^2 ( y )
  +\delta (x-y)\right\}\nonumber\\
   &+&  \left(\begin{array}{c}
    x \leftrightarrow \bar{x}\\
    y \leftrightarrow \bar{y}
  \end{array}\right)\label{EKSSNLO}
\end{eqnarray}
The corresponding evolution kernels in the case of forward scattering have
the following form
\begin{eqnarray}
 {\mbox {\bf P}}^{(0)}(x,y)&=&
 \left(\begin{array}{cc}
  2\bar z-\delta(1-z)&2  \\[3mm]
  2 z&-\delta(1-z)
  \end{array}\right)
\end{eqnarray}
\begin{eqnarray}
{\mbox {\bf P}}^{(1)}_{\!\!\psi \psi}(z)&=&-7\bar z-2\bar z\ln(\bar z)+\bar z\ln (z)+(1+z)\ln^2(z)+\delta(1-z)\\
{\mbox {\bf P}}^{(1)}_{\!\!\varphi \psi}(z)&=&-7+ z-2\ln(\bar z)+(1-2z)\ln (z)+\ln^2(z)\\
{\mbox {\bf P}}^{(1)}_{\!\!\psi \varphi}(z)&=&1-7 z-2 z\ln(\bar z)+(z-2)\ln (z)-z\ln^2(z)\\
{\mbox {\bf P}}^{(1)}_{\!\!\varphi \varphi}(z)&=&\bar z(1-2\ln(z))+\delta(1-z)
\end{eqnarray}

\subsubsection{Fermionic operator}

The calculation of evolution kernel for fermionic DA is similar to the calculation of evolution
kernels for bosonic DA. However, in this case there is one subtlety worth to mention. It is
necessary to keep track of what are the meanings of $x$ and $y$ in the expression for evolution
kernels.

\begin{figure}[ht]
\vspace{-1.5cm}
\begin{center}
\begin{picture}(680,120)(0,0)
\SetScale{0.65}
\SetWidth{1}
\Line(105,109)(96,91)
\Line(104,91)(95,109)
\Line(40,-20)(58,15)
\DashLine(58,15)(78,55){7}
\Line(78,55)(105,109)
\DashLine(160,-20)(142,15){7}
\Line(142,15)(122,55)
\DashLine(122,55)(104,91){7}
\CArc(100,100)(10,0,360)
\Line(122,55)(78,55)
\Line(142,15)(58,15)
\Vertex(122,55){3}
\Vertex(78,55){3}
\Vertex(142,15){3}
\Vertex(58,15){3}
\Text(65,-25)[c]{(a)}

\Line(255,109)(246,91)
\Line(254,91)(245,109)
\DashLine(190,-20)(208,15){7}
\Line(208,15)(255,109)
\Line(310,-20)(292,15)
\DashLine(292,15)(254,91){7}
\CArc(250,100)(10,0,360)
\DashCArc(228,55)(20,248,61){7}
\Line(292,15)(208,15)
\Vertex(236,72){3}
\Vertex(219,38){3}
\Vertex(292,15){3}
\Vertex(208,15){3}
\Text(165,-25)[c]{(b)}

\Line(405,109)(396,91)
\Line(404,91)(395,109)
\DashLine(340,-20)(358,15){7}
\Line(358,15)(405,109)
\Line(460,-20)(442,15)
\DashLine(442,15)(404,91){7}
\CArc(400,100)(10,0,360)
\DashCArc(400,15)(20,180,360){7}
\Line(442,15)(420,15)
\Line(442,15)(358,15)
\Line(380,15)(358,15)
\Vertex(420,15){3}
\Vertex(380,15){3}
\Vertex(442,15){3}
\Vertex(358,15){3}
\Text(262,-25)[c]{(c)}

\DashLine(555,109)(536,72){7}
\Line(555,109)(546,91)
\Line(554,91)(545,109)
\DashLine(508,15)(519,38){7}
\Line(490,-20)(508,15)
\Line(592,15)(545,109)
\DashLine(610,-20)(592,15){7}
\CArc(550,100)(10,0,360)
\CArc(528,55)(20,0,360)
\Line(592,15)(508,15)
\Vertex(536,72){3}
\Vertex(519,38){3}
\Vertex(592,15){3}
\Vertex(508,15){3}
\Text(360,-25)[c]{(d)}

\Line(705,109)(658,15)
\Line(705,109)(696,91)
\Line(704,91)(695,109)
\Line(640,-20)(658,15)
\DashLine(742,15)(704,91){7}
\DashLine(760,-20)(742,15){7}
\CArc(700,100)(10,0,360)
\DashLine(742,15)(658,15){7}
\DashLine(742,15)(678,55){7}
\Vertex(678,55){3}
\Vertex(742,15){3}
\Vertex(658,15){3}
\Text(457,-25)[c]{(e)}
\end{picture}
\end{center}
\vspace{10mm}
\caption{Two-loop diagrams for operator Eq.~(\ref{ConfOpFWZ1}) with fermion quantum numbers}
\label{TwoLoopFermion}
\end{figure}
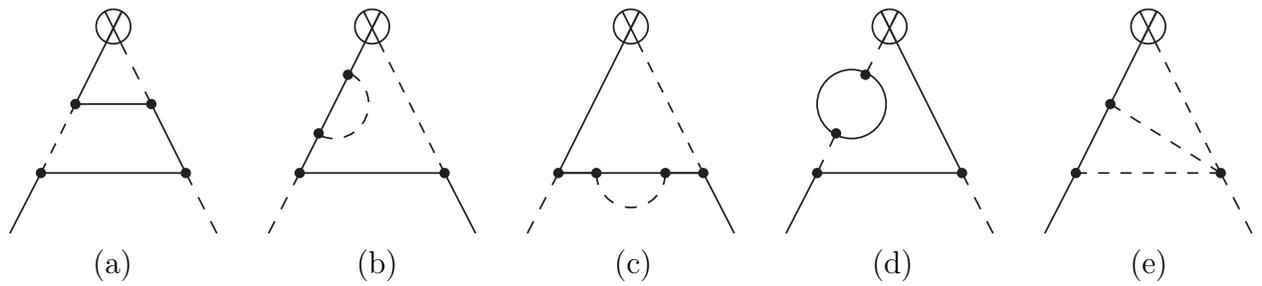

Now, for example if we agreed that $x$ and $y$ are the fractions of momentum for scalar field in
fermionic operator, then we have the following answers for evolution kernels in leading
and next-to-leading orders
(see diagrams in Fig.~\ref{OneLoop}.e and Fig.~\ref{TwoLoopFermion})
\begin{eqnarray}
   {\mbox {\bf V}}_{\!\!{\bf {fer}}}^{(0)}(x,y)&=& \theta(x<\bar y)
   \left\{-\frac{2}{\bar y}\right\}-\delta(x-y)\label{EKFerLO}\\
   {\mbox {\bf V}}_{\!\!{\bf {fer}}}^{(1)}(x,y)&=&
   \displaystyle \theta ( x < y )
   \left\{2\frac{\ln(x)\ln(y)}{\bar y} -
   \frac{\ln^2(y)}{\bar y}
   \right\} +
   \theta ( x > y )
   \left\{\frac{\ln^2(x)}{\bar y}\right\}\nonumber\\
  &+&\frac{1}{\bar y}\ln(x)- \theta ( x > \bar y )
   \left\{\frac{1}{\bar y}\ln\left(1-\frac{\bar x}{y}\right)\right\}\nonumber\\
  &+&\theta ( x < \bar y )\left\{
   \frac{6}{\bar y} +
   \frac{1}{\bar y}\ln\left(1-\frac{x}{\bar y}\right) +
   \frac{1}{\bar y}\ln\left(\frac{x}{\bar y}\right)\right\}+\delta(x-y)\label{EKFerNLO}
\end{eqnarray}
Evolution kernel in the case of forward scattering has the following form
\begin{eqnarray}
{\mbox {\bf P}}^{(0)}_{{\mathbf {fer}}}(z)&=&2-\delta(1-z)\\
{\mbox {\bf P}}^{(1)}_{{\mathbf {fer}}}(z)&=&-6-\ln(z)-2\ln(\bar z)+\ln^2(z)+\delta(1-z)
\end{eqnarray}

\subsection{Polarized case}

In polarized case the results for each contributing diagram differ from those in
unpolarized case only by overall sign.

\subsubsection{Non-singlet case}

In non-singlet case the diagrams Fig.~\ref{OneLoop}.a and Fig.~\ref{TwoLoopQQ}.b 
change the overall sign and thus the final result for the non-singlet evolution kernel in polarized case
has the following form:
\begin{eqnarray}
{\widetilde{\mbox {\bf V}}}_{\!\! NS}(x,y)&=&\frac{\alpha}{4 \pi}
{\widetilde{\mbox {\bf V}}}^{(0)}_{\!\! NS}(x,y)+
\left(\frac{\alpha}{4 \pi}\right)^2{\widetilde{\mbox {\bf V}}}^{(1)}_{\!\! NS}(x,y)\\
{\widetilde{\mbox {\bf V}}}^{(0)}_{\!\! NS}(x,y)&=& \theta(x<y)\left\{-2\frac{x}{y}-\delta(x-y)\right\}+
  \left(\begin{array}{c}
    x \leftrightarrow \bar{x}\\
    y \leftrightarrow \bar{y}
  \end{array}\right)\\
{\widetilde{\mbox {\bf V}}}^{(1)}_{\!\! NS}(x,y)&=& \theta(x<y)\left\{
4 \frac{x}{y}
+ \frac{\bar{x}}{\bar{y}} \ln ( \bar{x} )
+ \frac{x}{y} \ln ( x )
- \frac{1}{\bar y}\left(1-\frac{{x}}{{y}}\right) \ln \left( 1 - \frac{x}{y} \right) \right. \nonumber\\
&+& \left.
\frac{\bar{x}}{y} \ln^2 ( \bar x )
+ 2 \frac{x}{\bar{y}} \ln ( x ) \ln ( y  )
- \frac{x}{\bar{y}} \ln^2 ( y )
+\delta(x-y)\right\}+
  \left(\begin{array}{c}
    x \leftrightarrow \bar{x}\\
    y \leftrightarrow \bar{y}
  \end{array}\right)\label{KernelNSPol}
\end{eqnarray}
In the forward case the evolution kernels are given by
\begin{eqnarray}
\widetilde{\mbox {\bf P}}^{(0)}_{NS}(z)&=&-2\bar z-\delta(1-z)\\
\widetilde{\mbox {\bf P}}^{(1)}_{NS}(z)&=&4\bar z+2\bar z\ln(\bar z)+2\bar z\ln (z)+(1+z)\ln^2(z)+\delta(1-z)
\end{eqnarray}

\subsubsection{Singlet case}

Changing the overall sign of diagrams in the cases where it is necessary and
taking into account mixing between operators (\ref{qqp}) and (\ref{ssp})
the evolution kernel in the singlet case has the following form
\begin{eqnarray}
 \widetilde{\mbox {\bf V}}(x,y)&=&
 \left(\begin{array}{cc}
  \widetilde{\mbox{\bf V}}_{\!\!\psi \psi}(x,y)&\widetilde{\mbox{\bf V}}_{\!\!\psi \varphi}(x,y)  \\[3mm]
  \widetilde{\mbox{\bf V}}_{\!\!\varphi \psi}(x,y)&\widetilde{\mbox{\bf V}}_{\!\!\varphi \varphi}(x,y)
  \end{array}\right)
\end{eqnarray}
with the elements of this matrix in leading order given by
\begin{eqnarray}
  \widetilde{\mbox{\bf V}}^{(0)}_{\!\!\psi \psi}(x,y) & = &\theta(x<y)\left\{
  -2\frac{x}{y} -\delta (x-y)\right\}
  +  \left(\begin{array}{c}
    x \leftrightarrow \bar{x}\\
    y \leftrightarrow \bar{y}
  \end{array}\right)\nonumber\\
  \widetilde{\mbox {\bf V}}^{(0)}_{\!\!\psi \varphi}(x,y) & = &\theta(x<y)
  \Bigl\{-x \Bigr\}
    -  \left(\begin{array}{c}
    x \leftrightarrow \bar{x}\\
    y \leftrightarrow \bar{y}
  \end{array}\right)\nonumber\\
  \widetilde{\mbox {\bf V}}^{(0)}_{\!\!\varphi \psi}(x,y) & = &\theta(x<y)\left\{
  \frac{1}{y}\right\}
    -  \left(\begin{array}{c}
    x \leftrightarrow \bar{x}\\
    y \leftrightarrow \bar{y}
  \end{array}\right)\nonumber\\
  \widetilde{\mbox {\bf V}}^{(0)}_{\!\!\varphi \varphi}(x,y) & = &\theta(x<y)
  \Bigl\{2 -\delta (x-y)\Bigr\}
   +  \left(\begin{array}{c}
    x \leftrightarrow \bar{x}\\
    y \leftrightarrow \bar{y}
  \end{array}\right)
\end{eqnarray}
and in next-to-leading order
\begin{eqnarray}
 \widetilde {\mbox{\bf V}}^{(1)}_{\!\!\psi \psi}(x,y) & = &\theta(x<y)\left\{
   6 \frac{x}{y}
  +\frac{x}{y} \ln \left( \frac{x}{y} \right)
  + \frac{\bar{x}}{\bar{y}} \ln ( \bar{x} )
  + \frac{x}{y} \ln ( x )
  - \frac{1}{\bar y}\left(1-\frac{x}{y}\right) \ln \left( 1 - \frac{x}{y} \right) \right.\nonumber\\
  &+&\left. 
   \frac{\bar{x}}{y} \ln^2 ( \bar{x} )
  + 2 \frac{x}{\bar{y}} \ln ( x ) \ln ( y )
  - \frac{x}{\bar{y}} \ln^2 ( y )
  +\delta (x-y)\right \}
+  \left(\begin{array}{c}
    x \leftrightarrow \bar{x}\\
    y \leftrightarrow \bar{y}
  \end{array}\right)\\
 \widetilde {\mbox {\bf V}}^{(1)}_{\!\!\psi \varphi}(x,y) & = &\theta(x<y)\left\{
  6 x
  + x \ln \left( \frac{x}{y} \right)
  + \bar{x} \ln ( \bar{x} )
  - x \ln ( x )
  + \ln \left( 1 - \frac{x}{y} \right)\right. \nonumber\\
  &+& \left.\bar{x} \ln^2 ( \bar{x} )
  - 2 x \ln ( x ) \ln ( y )
  + x \ln^2 ( y )\right.\biggr\}
    -  \left(\begin{array}{c}
    x \leftrightarrow \bar{x}\\
    y \leftrightarrow \bar{y}
  \end{array}\right)\\
 \widetilde {\mbox {\bf V}}^{(1)}_{\!\!\varphi \psi}(x,y) & = &\theta(x<y)\left\{
  -\frac{6}{y}
  - \frac{1}{y} \ln \left( \frac{x}{y} \right)
  + \frac{1}{\bar{y}} \ln ( \bar{x} )
  - \frac{1}{y} \ln ( x )
  - \frac{1}{y\bar{y}} \ln \left( 1 - \frac{x}{y} \right) \right.\nonumber\\
  &+&\left. 
   \frac{1}{y} \ln^2 ( \bar{x} )
  - \frac{2}{\bar{y}} \ln ( x ) \ln ( y )
  +  \frac{1}{\bar{y}} \ln^2 ( y )\right\}
    -  \left(\begin{array}{c}
    x \leftrightarrow \bar{x}\\
    y \leftrightarrow \bar{y}
  \end{array}\right)\\
  \widetilde{\mbox {\bf V}}_{\!\!\varphi \varphi}(x,y) & = &\theta(x<y)\left\{
   -2
   - \ln \left( \frac{x}{y} \right)
   + \ln^2 (\bar{x} )
   + 2 \ln ( x ) \ln ( y )
   - \ln^2 ( y )+\delta (x-y)\right\}\nonumber\\
   &+&  \left(\begin{array}{c}
    x \leftrightarrow \bar{x}\\
    y \leftrightarrow \bar{y}
  \end{array}\right)
\end{eqnarray}
The corresponding kernels in the case of forward scattering are found to be
\begin{eqnarray}
 \widetilde{\mbox {\bf P}}^{(0)}(x,y)&=&
 \left(\begin{array}{cc}
  -2\bar z-\delta(1-z)&-2  \\[3mm]
  -2 z&-\delta(1-z)
  \end{array}\right)
\end{eqnarray}
\begin{eqnarray}
\widetilde{\mbox {\bf P}}^{(1)}_{\!\!\psi \psi}(z)&=&
7\bar z+2\bar z\ln(\bar z)+\bar z\ln (z)+(1+z)\ln^2(z)+\delta(1-z)\\
\widetilde{\mbox {\bf P}}^{(1)}_{\!\!\varphi \psi}(z)&=&
7-z+2\ln(\bar z)+(3-2z)\ln (z)+\ln^2(z)\\
\widetilde{\mbox {\bf P}}^{(1)}_{\!\!\psi \varphi}(z)&=&
-1+7 z+2 z\ln(\bar z)+(3z-2)\ln (z)-z\ln^2(z)\\
\widetilde{\mbox {\bf P}}^{(1)}_{\!\!\varphi \varphi}(z)&=&
-\bar z(1+2\ln(z))+\delta(1-z)
\end{eqnarray}

\subsubsection{Fermionic operator}
As it can be easily seen from Eqs.~(\ref{ConfOpFWZ1}) and (\ref{ConfOpFWZ2})
there is a simple relation between fermionic operators in polarized and unpolarized cases
\begin{equation}
{\widetilde{\mathcal O}}^{\mathbf {fer}}_{j,l}=
-\gamma_5{{\mathcal O}}^{\mathbf {fer}}_{j,l}
\end{equation}
Thus, the evolution kernel for fermionic DA in polarized case differs only by
overall sign from unpolarized one.

\section{Conformal Ward Identity approach}

As was shown in Refs.~\cite{MPhi3,MQEDNS}
 non-diagonal part of anomalous dimension matrix may be obtained
by studying renormalized Conformal Ward Identities (CWI) for the Green functions of elementary
fields with the insertions of corresponding conformal operators. To deduce the
necessary identities, one starts with the generating functional for renormalized
disconnected Green functions of conformal operators. Exploiting its invariance
with respect to transformations of collinear conformal group we get
\begin{equation}
\langle[{\cal O}_{jl}](\delta^G\chi)\rangle=
-\langle[(\delta^G{\cal O}_{jl}])\chi\rangle
-\langle[{\cal O}_{jl}](\delta^G[S])\chi\rangle\,,\label{Equality}
\end{equation}
where $<A>$ denotes averaging over the vacuum of the time ordered product
$T A \exp(i[S])$ and $\chi=\prod_i \phi_i$ stands for the product of elementary
fields at different space-time points. The left hand side of the above equality is the
differential operator acting on renormalized Green function
$\langle[{\cal O}_{jl}]\chi)\rangle$, which thus should be finite. Hence, the right
hand side of the equality is also finite. Using this fact we may determine the scale and
special conformal anomalous matrices for conformal operators under consideration.

Now, the crucial point is to note, that scale and special conformal anomalies are in fact
related. The relation between matrices of scale and special conformal anomalies could be
obtained by considering the action of the commutator of special conformal transformation with
dilation $[\delta^D,\delta^K_{-}] = \delta^K_{-}$ on Green function with conformal operator insertion.
As a result we obtain the following relation
\begin{equation}
\left[\hat a(l)+\hat\gamma^c(\alpha,l)+\frac{\beta(\alpha)}{\alpha}\hat b(l),
\hat\gamma(\alpha)\right]=0\,,\label{FinalEq}
\end{equation}
where $\hat\gamma^c(\alpha,l)$ is matrix of special conformal anomalies,
$\hat a(l)$, $\hat b(l)$ is some matrix coefficients and $\hat \gamma(\alpha)$ are
the anomalous dimensions matrix, we want to find. Due to the fact, that matrices
$\hat a(l)$ and $\hat b(l)$ do not depend on coupling constant (in contrast with all
other quantities, appearing in Eq.~(\ref{FinalEq})), it is possible
to solve this equation recursively at a given order of perturbation theory and find
the relation between non-diagonal elements of anomalous dimension matrix
$\gamma^{{\mathrm {ND}}}_{jk}$ in $n^{th}$-order and matrix of special
conformal anomalies $\hat\gamma^c(\alpha,l)$ in $(n-1)$-order of perturbation
theory. For example, at LO we have $\gamma^{{\mathrm {ND}}}_{jk}=\delta_{jk}\gamma^{{\mathrm {D}}}_{k}$.
It should be noted, that in the absence of gauge field the final expression is considerably
simplified and one only needs to know the matrices $a(l)$ £ $b(l)$. More details about this method
in the case of fermionic operator could be found in Appendix C. Below we present the results,
obtained within this framework, for non-diagonal parts of anomalous dimensions matrix of conformal
operators, we are interested in.

\subsection{Unpolarized case}

\subsubsection{Non-singlet case}
Consider first the case of non-singlet bosonic operator. Solving Eq.~(\ref{FinalEq})
in the next-to-leading order, it is easy to find the following general expression
for non-diagonal anomalous dimensions matrix \cite{MPhi3,MQEDNS}
\begin{eqnarray}
{}^{\psi\psi}\!\gamma^{\mathrm {ND}}_{j,k}&=&
\left({}^{\psi\psi}\!\gamma^{(0)}_{j}-{}^{\psi\psi}\!\gamma^{(0)}_{k}\right)d_{j,k}
\left(b_0+\gamma_\Phi-{}^{\psi\psi}\!\gamma^{(0)}_{k}\right)\,,\label{ND0}
\end{eqnarray}
where
\be
{}^{\psi\psi}\!\gamma^{(0)}_{n}
=\frac{2}{n+1}-\frac{2}{n+2}-1 \nonumber
\ee
is the corresponding leading order anomalous dimension of operator (\ref{qqn}) in forward case and
\begin{equation}
d_{j,k}=
\left\lbrace
\begin{array}{cc}
\displaystyle\frac{2k+3}{(j+k+3)(j-k)}\quad\mathrm{if}&j-k> 0\;\;\;\mathrm{and\;\;even}\\
&\\
0&\mathrm{otherwise}
\end{array}\right.
\end{equation}
Taking into account the next-to-leading order anomalous dimension in forward case
\begin{eqnarray}
^{\psi\psi}\gamma^{(1)}_n&=&\frac{2}{(n+1)^3} - \frac{2}{(n+1)^2} - \frac{4}{n+1}
+ \frac{2}{( n+2 )^3} +
 \frac{2}{( n+2 )^2} + \frac{4}{n+2}\nonumber\\
&+& \frac{2\,S_1(n+1)}{n+1}- \frac{2\,S_1(n+2)}{n+2}+1
\end{eqnarray}
it not hard to verify that the relation~(\ref{EKADRel}) holds true.

\subsubsection{Singlet case}

In the singlet case the expression for non-diagonal anomalous dimensions matrix should be
modified to take into account operator mixing. In the case of Wess-Zumino model the
corresponding expressions are \cite{MBQEDS}
\begin{eqnarray}
{}^{\psi\psi}\!\gamma^{\mathrm {ND}}_{j,k}&=&
\left({}^{\psi\psi}\!\gamma^{(0)}_{j}-{}^{\psi\psi}\!\gamma^{(0)}_{k}\right)d_{j,k}
\left(b_0-{}^{\psi\psi}\!\gamma^{(0)}_{k}\right)
-\left({}^{\psi\phi}\!\gamma^{(0)}_{j}-{}^{\psi\phi}\!\gamma^{(0)}_{k}\right)d_{j,k}
{}^{\phi\psi}\!\gamma^{(0)}_{k}\label{ND1}\\
{}^{\psi\phi}\!\gamma^{\mathrm {ND}}_{j,k}&=&
\left({}^{\psi\phi}\!\gamma^{(0)}_{j}-{}^{\psi\phi}\!\gamma^{(0)}_{k}\right)d_{j,k}
\left(b_0-{}^{\phi\phi}\!\gamma^{(0)}_{k}\right)
-\left({}^{\psi\psi}\!\gamma^{(0)}_{j}-{}^{\psi\psi}\!\gamma^{(0)}_{k}\right)d_{j,k}
{}^{\psi\phi}\!\gamma^{(0)}_{k}\label{ND2}\\
{}^{\phi\psi}\!\gamma^{\mathrm {ND}}_{j,k}&=&
\left({}^{\phi\psi}\!\gamma^{(0)}_{j}-{}^{\phi\psi}\!\gamma^{(0)}_{k}\right)d_{j,k}
\left(b_0-{}^{\psi\psi}\!\gamma^{(0)}_{k}\right)
-\left({}^{\phi\phi}\!\gamma^{(0)}_{j}-{}^{\phi\phi}\!\gamma^{(0)}_{k}\right)d_{j,k}
{}^{\phi\psi}\!\gamma^{(0)}_{k}\label{ND3}\\
{}^{\phi\phi}\!\gamma^{\mathrm {ND}}_{j,k}&=&
\left({}^{\phi\phi}\!\gamma^{(0)}_{j}-{}^{\phi\phi}\!\gamma^{(0)}_{k}\right)d_{j,k}
\left(b_0-{}^{\phi\phi}\!\gamma^{(0)}_{k}\right)
-\left({}^{\phi\psi}\!\gamma^{(0)}_{j}-{}^{\phi\psi}\!\gamma^{(0)}_{k}\right)d_{j,k}
{}^{\psi\phi}\!\gamma^{(0)}_{k}\,,\label{ND4}
\end{eqnarray}
where ${}^{ab}\!\gamma^{(0)}_{j}$ are leading order anomalous dimensions of singlet
operators (\ref{qqn}) and (\ref{ssn}):
\begin{equation}
\Gamma_{W\!Z}=
\begin{array}{|cc|}
{}^{\psi\psi}\gamma_n
&
{}^{\psi\phi}\gamma_n
\\[5mm]
{}^{\phi\psi}\gamma_n
&
{}^{\phi\phi}\gamma_n
\end{array}=
\begin{array}{|cc|}
\displaystyle \frac{2}{n+1} - \frac{2}{n+2}-1&
\quad\displaystyle \frac{2}{n+1}\frac{n+1}{2}\quad\\[8mm]
\displaystyle \frac{2}{n+2}\frac{2}{n+1}&\quad-1\quad
\end{array}\label{MADMF1L}
\end{equation}
Note, that in our normalization of Gegenbauer polynomials, the anomalous dimensions
of conformal operators (\ref{qqn}) and (\ref{ssn}) in singlet case differ from
corresponding anomalous dimensions of the usual (non-conformal) Wilson twist-2 operators
appearing in the description of DIS by shift in argument $n\rightarrow n+1$ and
normalization factors. Here and in what follows we will show explicitly additional normalization factors
$\frac{2}{n+1}$ or $\frac{n+1}{2}$.
For example, our results for ${}^{\psi\phi}\gamma_n$ and ${}^{\phi\psi}\gamma_n$
in Eq.~(\ref{MADMF1L}) contain explicit multiplication by these factors.

The next-to-leading order diagonal elements of anomalous dimensions
matrix in this case are given by
\begin{eqnarray}
{}^{\psi\psi}\gamma^{(1)}_n
&=&\displaystyle \frac{2}{(n+1)^3} - \frac{1}{(n+1)^2} - \frac{7}{n+1}
+ \frac{2}{( n+2 )^3} -
 \frac{1}{( n+2 )^2} + \frac{7}{n+2}\nonumber\\
&+& \frac{2\,S_1(n+1)}{n+1}- \frac{2\,S_1(n+2)}{n+2}
+1\label{ADQQNLO}\\
{}^{\psi\phi}\gamma^{(1)}_n
&=&\displaystyle \left(\frac{2}{(n+1)^3} - \frac{1}{(n+1)^2}
- \frac{7}{n+1} + \frac{2}{(n+2)^2}
+ \frac{1}{n+2} + \frac{2\,{S_1(n+1)}}{n+1}\right)\frac{n+1}{2}\label{ADQSNLO}\\
{}^{\phi\psi}\gamma^{(1)}_n
&=&\displaystyle \left(  \frac{2}{(n+1)^2} + \frac{1}{n+1} - \frac{2}{(n+2)^3}
  + \frac{1}{(n+2)^2} - \frac{7}{n+2} +
   \frac{2\,S_1(n+1)}{n+2}\right)\frac{2}{n+1}\label{ADSQNLO}\\
{}^{\phi\phi}\gamma^{(1)}_n
&=&\displaystyle  \frac{2}{(n+1)^2} + \frac{1}{n+1}
   - \frac{2}{(n+2)^2} - \frac{1}{n+2}+1\label{ADSSNLO}
\end{eqnarray}

In singlet case the relation similar to Eq.~(\ref{EKADRel})
has the following matrix form
\begin{eqnarray}
\int^1_0\! &dx&
\left(\begin{array}{cc}
C^{3/2}_n(x-\bar x)&
C^{1/2}_{n+1}(x-\bar x)
\end{array}\right)
 \left(\begin{array}{cc}
  {\mbox{\bf V}}_{\!\!\psi \psi}(x,y)&{\mbox{\bf V}}_{\!\!\psi \varphi}(x,y) \nonumber \\[5mm]
  {\mbox{\bf V}}_{\!\!\varphi \psi}(x,y)&{\mbox{\bf V}}_{\!\!\varphi \varphi}(x,y)
  \end{array}\right)=\\
=&\displaystyle
&\sum^n_{k=0}
\left(\begin{array}{cc}
{}^{\psi\psi}\gamma_{n,k}
&
{}^{\psi\phi}\gamma_{n,k}
\\[5mm]
{}^{\phi\psi}\gamma_{n,k}
&
{}^{\phi\phi}\gamma_{n,k}
\end{array}\right)
\left(\begin{array}{c}
C^{3/2}_k(y-\bar y)
\\[5mm]
C^{1/2}_{k+1}(y-\bar y)
\end{array}\right)
\end{eqnarray}
It could be easily verified substituting explicit expressions for NLO evolution
kernels Eqs.~(\ref{EKQQNLO}-\ref{EKSSNLO}) and NLO anomalous dimensions Eqs.~(\ref{ADQQNLO}-\ref{ADSSNLO}).

\subsubsection{Fermionic operator}

The expression for non-diagonal anomalous dimensions matrix of operators with fermionic quantum
numbers (\ref{ConfOpFWZ1}) could be derived similar to the case of bosonic operators (for more
details see Appendix C). It is given by
\begin{eqnarray}
{}^{\bf {fer}}\!\gamma^{\mathrm {ND}}_{j,k}&=&
\left({}^{\bf {fer}}\!\gamma^{(0)}_{j}-{}^{\bf {fer}}\!\gamma^{(0)}_{k}\right)d_{j,k}^{\bf{fer}}
\left(\beta_0-{}^{\bf {fer}}\!\gamma^{(0)}_{k}\right)\,,\label{NDF}
\end{eqnarray}
where
\begin{equation}
{}^{\bf {fer}}\!\gamma^{(0)}_{n} =(-1)^{n+1}\frac{2}{n+1}-1\label{AD1LFer}
\end{equation}
is the corresponding leading order anomalous dimension of fermionic operator
(\ref{ConfOpFWZ1}) in forward case. The matrix $d_{j,k}^{\bf{fer}}$ is defined
through the derivative of corresponding Jacobi polynomial over its
indices:
\begin{equation}
\left.\frac{d}{d \nu}P^{(1+\nu,\nu)}_j(z)\right|_{\nu=0}=2
\sum^j_{k=0}d_{j,k}^{\bf{fer}}P^{(1,0)}_k(z)
\end{equation}
The explicit expression for $d_{j,k}^{\bf{fer}}$ ($j>k$) has the following form
\begin{equation}\label{djkfer}
d_{j,k}^{\bf{fer}}=
(-1)^{j-k}\displaystyle\frac{1}{j-k}\frac{(j+1)+(-1)^{j-k}(k+1)}{j+k+2}\frac{k+1}{j+1}
\end{equation}
Taking into account the next-to-leading order anomalous dimension of this operator in forward case
\begin{equation}
{}^{\bf{fer}}\gamma^{(1)}_n=\frac{2}{(n+1)^3}+1+(-1)^{n+1}\left(\frac{1}{(n+1)^2}-
\frac{6}{n+1}+\frac{S_1(n+1)}{n+1}\right)\label{AD2LFer}
\end{equation}
it easy to verify, that the relation similar to Eq.~(\ref{EKADRel})
\begin{equation}
\int^1_0 d xP^{(1,0)}_j(x-\bar x){\mbox {\bf V}}_{\bf{fer}}(x,y|\alpha)=
\sum_{k=0}^j{}^{\bf{fer}}\gamma_{jk}(\alpha)P^{(1,0)}_k(y-\bar y)\label{EKADRelFer}
\end{equation}
is valid.

\subsection{Polarized case}

In the polarized case non-diagonal anomalous dimensions matrices have the same expression as
in unpolarized case (see. Eqs.~(\ref{ND0}),~(\ref{ND1})-(\ref{ND4}) and~(\ref{NDF})).
The only change, which should be done, is the replacement of unpolarized forward anomalous dimensions
$\gamma_j$ with polarized ones $\tilde \gamma_j$.

\subsubsection{Non-singlet case}

Taking into account the change of sign in diagrams in
Fig.~\ref{OneLoop}.a and Fig.~\ref{TwoLoopQQ}.b we find that
polarized leading order anomalous dimension is given by
\begin{equation}
{}^{\psi\psi}\!\tilde\gamma^{(0)}_{n} =-\frac{2}{n+1}+\frac{2}{n+2}-1 \,,
\end{equation}
whereas in next-to-leading order we have
\begin{eqnarray}
^{\psi\psi}{\tilde\gamma^{(1)}}_n&=&\frac{2}{(n+1)^3} - \frac{2}{(n+1)^2} + \frac{4}{n+1}
+ \frac{2}{( n+2 )^3} +
 \frac{2}{( n+2 )^2} - \frac{4}{n+2}\nonumber\\
&-& \frac{2\,S_1(n+1)}{n+1}
+ \frac{2\,S_1(n+2)}{n+2}+1
\end{eqnarray}
It easy to see , that the relation~(\ref{EKADRel}) holds true.
\subsubsection{Singlet case}
In singlet case the leading order anomalous dimensions are
\begin{equation}
\widetilde\Gamma_{W\!Z}=
\begin{array}{|cc|}
{}^{\psi\psi}\tilde\gamma_n
&
{}^{\psi\phi}\tilde\gamma_n
\\[5mm]
{}^{\phi\psi}\tilde\gamma_n
&
{}^{\phi\phi}\tilde\gamma_n
\end{array}=
\begin{array}{|cc|}
\displaystyle -\frac{2}{n+1} + \frac{2}{n+2}-1&
\quad\displaystyle -\frac{2}{n+1}\frac{n+1}{2}\quad\\[8mm]
\displaystyle -\frac{2}{n+2}\frac{2}{n+1}&\quad-1\quad
\end{array}\label{MADMFP1L}
\end{equation}
In the next-to leading order the diagonal elements of the anomalous dimensions matrix
have the following expressions
\begin{eqnarray}
{}^{\psi\psi}\tilde\gamma^{(1)}_n
&=&\displaystyle \frac{2}{(n+1)^3} - \frac{3}{(n+1)^2} + \frac{7}{n+1}
+ \frac{2}{( n+2 )^3} + \frac{3}{( n+2 )^2} - \frac{7}{n+2}\nonumber\\
&-& \frac{2\,S_1(n+1)}{n+1}+ \frac{2\,S_1(n+2)}{n+2}
+1\label{ADQQNLOP}\\
{}^{\psi\phi}\tilde\gamma^{(1)}_n
&=&\displaystyle \left(\frac{2}{(n+1)^3} - \frac{3}{(n+1)^2}
+ \frac{7}{n+1} + \frac{2}{(n+2)^2}
- \frac{1}{n+2} - \frac{2\,{S_1(n+1)}}{n+1}\right)\frac{n+1}{2}\label{ADQSNLOP}\\
{}^{\phi\psi}\tilde\gamma^{(1)}_n
&=&\displaystyle \left( \frac{2}{(n+1)^2} - \frac{1}{n+1} - \frac{2}{(n+2)^3}
  - \frac{3}{(n+2)^2} + \frac{7}{n+2}
  - \frac{2\,S_1(n+1)}{n+2}\right)\frac{2}{n+1}\label{ADSQNLOP}\\
{}^{\phi\phi}\tilde\gamma^{(1)}_n
&=&\displaystyle \frac{2}{(n+1)^2} - \frac{1}{n+1}
   - \frac{2}{(n+2)^2} + \frac{1}{n+2}+1\label{ADSSNLOP}
\end{eqnarray}

\section{Supersymmetric Ward Identity}

Now, let us recall, that all operators considered above form the representation
of $N=1$ supersymmetry algebra. Similar to CWI we may consider the supersymmetric Ward
identity~\cite{SUSYCWI} (its derivation goes along the same lines as the derivation of CWI).
As we will see in a moment, there is a host of remarkable consequences
for anomalous dimensions of our operators following from supersymmetric Ward
identity. To begin with, let us introduce the following combinations of conformal operators
for unpolarized (Eqs.~(\ref{qqn}), (\ref{ssn}) and (\ref{ConfOpFWZ1}))
and polarized (Eqs.~(\ref{qqp}), (\ref{ssp}) and (\ref{ConfOpFWZ2})) operators.
\begin{align}
{\mathcal S}^1_{j,l}
&=
\displaystyle\frac{2}{j+1}{\mathcal O}_{j,l}^{\psi}
+{\mathcal O}_{j,l}^{\phi}
&\qquad\qquad
{\mathcal P}^1_{j,l}
&=
\displaystyle\frac{2}{j+1}\widetilde{\mathcal O}_{j,l}^{\psi}
+\widetilde{\mathcal O}_{j,l}^{\phi}\label{scal1}\\
{\mathcal S}^2_{j,l}
&=
\displaystyle -\frac{j+1}{j+2}\;\frac{2}{j+1}{\mathcal O}_{j,l}^{\psi}
+{\mathcal O}_{j,l}^{\phi}
&
{\mathcal P}^2_{j,l}
&=
\displaystyle -\frac{j+1}{j+2}\;\frac{2}{j+1}\widetilde{\mathcal O}_{j,l}^{\psi}
+\widetilde{\mathcal O}_{j,l}^{\phi}\label{scal2}\\
{\mathcal V}_{j,l}
&=
{\mathcal O}_{j,l}^{\mathbf {fer}}
&
{\mathcal U}_{j,l}
&=
\widetilde{\mathcal O}_{j,l}^{\mathbf {fer}}\,,\label{ferm}
\end{align}
where the coefficients in front of operators define the matrix, diagonalizing
the matrices of forward anomalous dimensions (\ref{MADMF1L}) and~(\ref{MADMFP1L}).

Under restricted light-cone supersymmetry transformations the combinations of conformal operators introduced above transform as follows
\begin{eqnarray}
\delta^Q {\mathcal S}^1_{j,l}&=&(1-(-1)^j)\bar\epsilon {\mathcal V}_{j-1,l}^\dag\label{COs1}\\
\delta^Q {\mathcal S}^2_{j,l}&=&(1-(-1)^j)\bar\epsilon {\mathcal V}_{j,l}^\dag\label{COs2}\\
\delta^Q {\mathcal P}^1_{j,l}&=&(1+(-1)^j)\bar\epsilon {\mathcal
 V}_{j-1,l}^\dag\label{COp1}\\
\delta^Q {\mathcal P}^2_{j,l}&=&(1+(-1)^j)\bar\epsilon {\mathcal
 V}_{j,l}^\dag\label{COp2}\\
\delta^Q {\mathcal V}_{j,l}^\dag &=& i\partial_+\left(
{\mathcal S}^1_{j+1,l} +\gamma_5{\mathcal P}^2_{j,l}\right)\gamma_{-}\epsilon
+i\partial_+\left(
{\mathcal S}^2_{j,l} + \gamma_5{\mathcal P}^1_{j+1,l}\right)\gamma_{-}\epsilon\,.
\label{COfv}
\end{eqnarray}
The transformation low for the operators ${\cal U}_{j,k}$ foolowing from the observation
${\cal U}_{j,k}=-\gamma_5{\cal V}_{j,k}$ and the Eq.~(\ref{COp2}).
Now, we see that it is natural to combine bosonic and fermionic conformal operators into $N=1$
chiral supermultiplet.
\begin{equation}
\Phi=
\left({\mathcal S}^1+i{\mathcal P}^2\right)
+\sqrt{2}\theta  {\mathcal X}
+\theta\theta\left({\mathcal S}^2+i{\mathcal P}^1\right)\,,\label{SuperMult}
\end{equation}
where ${\mathcal V}_{j-1,l}$ is the Majorana fermion build
from the Weyl spinor ${\mathcal X}$.

It is easy to find, that the renormalized supersymmetric Ward identity in the regularization
scheme, preserving supersymmetry, has the following form~\cite{SUSYCWI}
($\mbox{\boldmath$\cal S$}_{j,l}$ denotes vector of operators  $S^1_{j,l}$ and $S^2_{j,l}$)
\begin{equation}
\langle [\mbox{\boldmath$\cal S$}_{jl}] \delta^Q {\cal X} \rangle
= - \langle \delta^Q [\mbox{\boldmath$\cal S$}_{jl}] {\cal X} \rangle
- \langle i [\mbox{\boldmath$\cal S$}_{jl}] (\delta^Q S) {\cal X} \rangle
\qquad\mbox{and}\quad
\langle \delta^Q [\mbox{\boldmath$\cal S$}_{jl}] {\cal X} \rangle = \mbox{finite}\,,\label{SCWI}
\end{equation}
where we used the fact, that renormalized action in supersymmetric regularization is invariant
with respect to supersymmetry transformations $\langle i [\mbox{\boldmath$\cal S$}_{jl}] (\delta^Q S) {\cal X} \rangle=0$.
The operators (\ref{COs1}) and (\ref{COs2}) mix under renormalization and thus we define
renormalized operators as (square brackets correspond to renormalized quantities)
\begin{equation}
\left[
\begin{array}{c}
{\cal S}^1\\ {\cal S}^2
\end{array}
\right]_{jl}
= \sum_{k = 0}^{j}
\left(\begin{array}{cc}
^{11}\!Z_{\cal S}& ^{12}\!Z_{\cal S}\\
^{21}\!Z_{\cal S}& ^{22}\!Z_{\cal S}
\end{array}\right)_{jk}
\left(\begin{array}{cc}
Z_{\phi}^{-1} & 0 \\
0 & Z_{\phi}^{-1}
\end{array}\right)
\left(
\begin{array}{c}
{\cal S}^1\\ {\cal S}^2
\end{array}
\right)_{kl}
\end{equation}
and as a consequence the renormalization group equation for these operators is given by
\begin{equation}
\frac{d}{d \ln \mu}
\left[
\begin{array}{c}
{\cal S}^1\\ {\cal S}^2
\end{array}
\right]_{jl}=
 \sum_{k = 0}^{j}
\left(\begin{array}{cc}
^{11}\!\gamma^{\cal S}& ^{12}\!\gamma^{\cal S}\\
^{21}\!\gamma^{\cal S}& ^{22}\!\gamma^{\cal S}
\end{array}\right)_{jk}
\left[
\begin{array}{c}
{\cal S}^1\\ {\cal S}^2
\end{array}
\right]_{kl}
\end{equation}
Now, from supersymmetric Ward identity (\ref{SCWI}) we get
($\sigma_k=\frac12(1-(-1)^k)$ and $Z_{jk} = 0$ for $k > j$)
\begin{equation}\label{RelAD}
\sum_{k = 0}^{j} \sum_{k' = 0}^{k}
\left(
\begin{array}{ll}
 {^{11}\!Z}_{\cal S}  &  {^{12}\!Z}_{\cal S}  \\
 {^{21}\!Z}_{\cal S}  &  {^{22}\!Z}_{\cal S}
\end{array}
\right)_{jk}
\sigma_k
\left(
\begin{array}{l}
\{ Z_{\cal V}^{-1} \}_{k - 1, k'}  \\
\{ Z_{\cal V}^{-1} \}_{k k'}
\end{array}
\right)
[{\cal V}_{k' l}]
= \mbox{finite}
\end{equation}
$1/\epsilon$ poles in (\ref{RelAD}) cancel, provided
\begin{eqnarray}
&&\sum_{k = 0}^{j} \left\{ {^{11}\!Z^{[1]}_{\cal S}} \right\}_{jk}
\, \sigma_k \, [{\cal V}_{k - 1, l}]
+ \sum_{k = 0}^{j} \left\{ {^{12}\!Z^{[1]}_{\cal S}} \right\}_{jk}
\, \sigma_k \, [{\cal V}_{kl}]
= \sigma_j \sum_{k = 0}^{j} \left\{ {Z^{[1]}_{\cal V}} \right\}_{j - 1, k}
[{\cal V} _{kl}] ,\\
&&\sum_{k = 0}^{j} \left\{ {^{21}\!Z^{[1]}_{\cal S}} \right\}_{jk}
\, \sigma_k \, [{\cal V} _{k - 1, l}]
+ \sum_{k = 0}^{j} \left\{ {^{22}\!Z^{[1]}_{\cal S}} \right\}_{jk}
\, \sigma_k \, [{\cal V} _{kl}]
= \sigma_j \sum_{k = 0}^{j} \left\{ {Z^{[1]}_{\cal V}} \right\}_{jk}
[{\cal V} _{kl}]
\end{eqnarray}
Noting, that ${\mathcal U}_{j,k}=-\gamma_5{\mathcal V}_{j,k}$, it is easy to derive analogous
relations for polarized operators ${\mathcal P}_{j,k}$.
Taking into account linear independence of fermionic operators $[{\cal V} _{kl}]$ we finally get
the following relations~\cite{SUSYCWI}
\begin{alignat}{6}\label{S11}
{}^{11}\!\gamma^{\cal S}_{2n + 1, 2m + 1}
&=& {^{22}\!\gamma}^{\cal P}_{2n, 2m} \ \ \;
&= {}^{\mathbf {fer}}\gamma_{2n, 2m} , &\qquad\quad m \leq n , \\
\label{S12}
{}^{12}\!\gamma^{\cal S}_{2n + 1, 2m + 1}
&=& {^{21}\!\gamma}^{\cal P}_{2n, 2m + 2} \
&= {}^{\mathbf {fer}}\gamma_{2n, 2m + 1} , &\qquad\quad  m \leq n - 1 , \\
\label{S21}
{^{21}\!\gamma}^{\cal S}_{2n + 1, 2m + 1}
&=& {^{12}\!\gamma}^{\cal P}_{2n + 2, 2m} \
&= {}^{\mathbf {fer}}\gamma_{2n + 1, 2m} , &\qquad\quad m \leq n \ , \\
\label{S22}
{^{22}\!\gamma}^{\cal S}_{2n + 1, 2m + 1}
&=& {^{11}\!\gamma}^{\cal P}_{2n + 2, 2m + 2}
&= {}^{\mathbf {fer}}\gamma_{2n + 1, 2m + 1} , &\qquad\quad m \leq n \ ,
\end{alignat}
and
\begin{eqnarray}
\label{Dok}
{^{12}\!\gamma}^{\cal S}_{2n + 1, 2n + 1}
&=& 0 \ , \quad
{^{12}\!\gamma}^{\cal P}_{2n, 2n} = 0 \ .
\end{eqnarray}
These relations allow us to find anomalous dimensions of bosonic operators
from known anomalous dimensions of fermionic operators and vice versa.
In particular, substituting explicit expression for non-diagonal anomalous dimension
of bosonic operators, one can determine matrix $d_{j,k}$ for fermionic
operators Eq.~(\ref{djkfer}). The found relations between anomalous dimensions of
conformal operators is special case of existing superconformal relations
for anomalous dimensions of conformal operators in supersymmetric theories.

To obtain from the results above the corresponding relations for anomalous dimensions
of conformal operators considered earlier, one will need the following transition formulae
among anomalous dimensions of conformal operators (\ref{scal1}) and (\ref{scal2})
and anomalous dimensions of operators (\ref{qqn}) and (\ref{ssn})
\begin{equation}
\left(
\begin{array}{r}
\displaystyle\frac{1}{k+1}\, {^{11}\!\gamma} \\[3mm]
\displaystyle\frac{1}{k+2}\, {^{12}\!\gamma} \\[3mm]
\displaystyle\frac{1}{k+1}\, {^{21}\!\gamma} \\[3mm]
\displaystyle\frac{1}{k+2}\, {^{22}\!\gamma}
\end{array}
\right)_{\!\!jk}
=\ \frac{1}{2k + 3}
\left(
\begin{array}{cccc}
\displaystyle\frac{k+2}{j+1}&\displaystyle\frac{k+2}{2}
&\displaystyle\frac{2}{j+1}  & 1 \\[3mm]
\displaystyle\frac{k+1}{j+1}  & \displaystyle\frac{k+1}{2}
& -\displaystyle\frac{2}{j+1}                  & -1\\[3mm]
\displaystyle\frac{k+2}{j+2}&-\displaystyle\frac{k+2}{2}
&  \displaystyle \frac{2}{j+2}     & -1  \\[3mm]
\displaystyle\frac{k+1}{j+2} &-\displaystyle\frac{k+1}{2}
& -\displaystyle\frac{2}{j+2}   & 1
\end{array}
\right)
\left(
\begin{array}{r}
^{\psi\psi}\gamma \\[5.4mm]
^{\phi\psi}\gamma \\[5.4mm]
^{\psi\phi}\gamma \\[5.4mm]
^{\phi\phi}\gamma
\end{array}
\right)_{\!\!jk}
\end{equation}
Now, let us explore some particular limits of equations
(\ref{S11})-(\ref{Dok}). From Eq.~(\ref{Dok}) it follows, that
(here and below $\gamma_{jj} \equiv \gamma_j$, $\gamma \equiv \gamma^V$,
$\widetilde\gamma \equiv \gamma^A$ and indexes $\cal S$ ($\cal P$) correspond to
unpolarized (polarized) cases):
\begin{equation}
\label{DokSUSY}
  {}^{\psi\psi}\gamma^i_j
+ \frac{j+1}{2}{}^{\phi\psi}\gamma^i_j
= \frac{2}{j+1}{}^{\psi\phi}\gamma^i_j
+ {}^{\phi\phi}\gamma^i_j ,
\quad i = {\cal S},{\cal P} ,
\end{equation}
where $j$ is {\it odd} for unpolarized case and {\it even} for polarized case.
This equation is analogous (up to redefinition of anomalous dimensions)
to the well-known Dokshitzer relation (last paper in Ref.~\cite{DGLAP}),
which was found first empirically in LO in the case of supersymmetric QCD
(the anomalous dimensions in supersymmetric QCD could be obtained from
those in ordinary QCD through substitutions $C_A=N_c$, $C_F=N_c$ and $N_f=\frac12 N_c$.).

From (\ref{S11}) and (\ref{S22}), we also have (for {\it odd} $j$)
\begin{alignat}{5}
 {}^{\psi\psi}\gamma^{\cal S}_{j}
+\frac{j+1}{2} {{}^{\phi\psi}\gamma}^{\cal S}_{j}
&=&
 {}^{\psi\psi}\gamma^{\cal P}_{j-1}
-\frac{2}{j}{{}^{\psi\phi}\gamma}^{\cal P}_{j-1}&={}^{\mathbf {fer}}\gamma_{j-1}&\,,\label{SUSY2a}\\
 {}^{\psi\psi}\gamma^{\cal P}_{j + 1}
+\frac{j+2}{2}{}^{\phi\psi}\gamma^{\cal P}_{j + 1}
&=&
 {}^{\psi\psi}\gamma^{\cal S}_j
-\frac{2}{j+1}{}^{\psi\phi}\gamma^{\cal S}_j&={}^{\mathbf {fer}}\gamma_j&. \label{SUSY2b}
\end{alignat}
Moreover, from equations (\ref{S12}) and (\ref{S21}) we get the following
relations between diagonal and non-diagonal elements of anomalous dimensions
matrices of conformal operators
\begin{alignat}{4}
\frac{2}{j+2}{{}^{\psi\phi}\gamma}_j^i
-\frac{j+1}{2}{{}^{\phi\psi}\gamma}_j^i
&=& \frac{j+1}{2j + 1} \Delta_{j + 1, j - 1}^k
&=&{}^{\mathbf {fer}}\gamma_{j,j-1}\,,
\label{for-nonfora}
\end{alignat}
where $j$ is {\it odd} for $i={\cal S}$ and $k={\cal P}$,
$j$ is {\it even} for $i={\cal P}$ and $k={\cal S}$ and
\begin{equation}
\Delta^i_{j + 1, j - 1}
\equiv
\frac{j}{j+2}{{}^{\psi\psi}\gamma}^i_{j + 1, j - 1}
\!\!+\frac{j}{2}{{}^{\phi\psi}\gamma}^i_{j + 1, j - 1}
\!\!-\frac{2}{j+2}{{}^{\psi\phi}\gamma}^i_{j + 1, j - 1}
\!\!-{{}^{\phi\phi}\gamma}^i_{j + 1, j - 1} .\label{Delta}
\end{equation}
In the leading order $\Delta^i_{j + 1, j - 1}$ is equal to zero, but starting
from next-to-leading order it acquires nonzero contribution.
Here, we would like to note, that all these relations are valid in all
orders of perturbation theory provided anomalous dimensions were evaluated
in supersymmetric scheme. In Wess-Zumino model such scheme could be identified
with the usual $\overline{\mathrm{MS}}$-scheme.

Next, the forward anomalous dimensions matrix in singlet case (\ref{MADMF1L}) is diagonalized
with the help of  matrix ${\mathbf D}$
\begin{equation}
{\mathbf D}=
\begin{array}{|cc|}
\displaystyle \frac{2}{j+1}&1\\[3mm]
\displaystyle -\frac{2}{j+2}&1
\end{array}\label{DiagMatr}
\end{equation}
so, that in leading order we have
\begin{equation}
\left[{\mathbf D}\Gamma_{\mathbf {WZ}}{\mathbf D}^{-1}\right]^{(0)}=
\begin{array}{|cc|}
\displaystyle \frac{2}{j+1}-1&0\\[3mm]
0&\displaystyle -\frac{2}{j+2}-1
\end{array}
\end{equation}
In next-to-leading order of perturbation theory the elements of forward
anomalous dimensions matrix of bosonic operators are given by (\ref{ADQQNLO})-(\ref{ADSSNLO}). Using
the same diagonalizing matrix $\mathbf D$ (\ref{DiagMatr}), as in leading order,
we get
\begin{eqnarray}
\left[{\mathbf D}\Gamma_{\mathbf {WZ}}{\mathbf D}^{-1}\right]^{(1)}\!\!&=&\label{MADMF2L}\\[5mm]
\hspace*{-30mm}&\hspace*{-30mm}&\hspace*{-40mm}
\begin{array}{|cc|}
\displaystyle \frac{2}{(j+1)^3} + \frac{1}{(j+1)^2} - \frac{6}{j+1} +
\frac{2\,S_1(j+1)}{j+1}+1 &0\\[5mm] \displaystyle \frac{4}{j+1}-\frac{4}{j+2}&
\!\!\!\!\!\!\!\!\!\!\displaystyle  \frac{2}{(j+2)^3}
- \frac{1}{(j+2)^2} + \frac{6}{j+2} - \frac{2\,S_1(j+2}{j+2}+1
\end{array}\nonumber
\end{eqnarray}

So, in leading order the following relations between elements of anomalous dimensions
matrix are valid
\begin{eqnarray}
{}^{\psi\psi}\gamma_j+\frac{j+1}{2}{}^{\phi\psi}\gamma_j&=&
\frac{2}{j+1}{}^{\psi\phi}\gamma_j+{}^{\phi\phi}\gamma_j\label{rel1}\\
{}^{\psi\psi}\gamma_j-\frac{j+1}{2}\frac{j+1}{j+2}{}^{\phi\psi}\gamma_j&=&
-\frac{2}{j+1}\frac{j+2}{j+1}{}^{\psi\phi}\gamma_j+{}^{\phi\phi}\gamma_j
\label{rel2}
\end{eqnarray}
However, in next-to-leading order the relation ~(\ref{rel2}) does not hold,
while the relation~(\ref{rel1}) is valid to all orders of perturbation theory.
Note, also, that from the relation~(\ref{rel1}) it follows that
\begin{equation}
{}^{\psi\psi}\gamma_j-\frac{2}{j+1}{}^{\psi\phi}\gamma_j
=
-\frac{j+1}{2}{}^{\phi\psi}\gamma_j+{}^{\phi\phi}\gamma_j\label{rel3}
\end{equation}
which is also valid to all orders in perturbation theory and in leading
order coincides with the relation~(\ref{rel2}). Moreover, modifying
the diagonalizing matrix (\ref{DiagMatr}) it is possible to show, that
the forward anomalous dimensions matrix will remain diagonal also in
next-to-leading order, and, what is more important, in all orders of perturbation
theory. Such matrix ${\mathbf {\widehat D}}$ is given by \cite{AK}
\footnote{We thank Anatoly Kotikov for fruitful discussions of this issue.}:
\begin{equation}
{\mathbf {\widehat D}}=
\begin{array}{|cc|}
\displaystyle\frac{2}{j+1}&1\\[3mm]
-\displaystyle\frac{j+1}{2}\frac{{}^{\phi\psi}\gamma_j}
{{}^{\psi\phi}\gamma_j}
&1
\end{array}\label{DiagMatrAll}\; ,
\end{equation}
In polarized case the forward anomalous dimensions matrix (\ref{MADMFP1L})
may be diagonalized with the help of the same matrix $\mathbf D$ (\ref{DiagMatr}),
as in unpolarized case and it is possible to find, that
\begin{equation}
\left[{\widetilde{\mathbf D}}{\widetilde\Gamma_{\mathbf {WZ}}}{\widetilde{\mathbf D}}^{-1}\right]^{(0)}=
\begin{array}{|cc|}
\displaystyle -\frac{2}{j+1}-1&0\\[3mm]
0&\displaystyle \frac{2}{j+2}-1\,.\label{MADWZ1LP}
\end{array}
\end{equation}
In next-to-leading order the elements of forward anomalous dimensions matrix in polarized
case are given by Eqs.~(\ref{ADQQNLOP})-(\ref{ADSSNLOP}).
Using the same diagonalization matrix, as in leading order, we get
\begin{eqnarray}
\left[{\widetilde{\mathbf D}}{\widetilde\Gamma}_{\mathbf {WZ}}
{\widetilde{\mathbf D}}^{-1}\right]^{(1)}\!\!&=&\label{MADMF2LP}\\[5mm]
\hspace*{-30mm}&\hspace*{-30mm}&\hspace*{-40mm} \begin{array}{|cc|}
\displaystyle -\frac{2}{(j+1)^3} + \frac{1}{(j+1)^2} - \frac{6}{j+1} +
\frac{2\,S_1(j+1)}{j+1}-1 &0\\[5mm]
\displaystyle \frac{4}{j+1}-\frac{4}{j+2}+\frac{8}{(j+2)^2}&
\!\!\!\!\!\!\!\!\!\!\displaystyle  -\frac{2}{(j+2)^3}
- \frac{1}{(j+2)^2} + \frac{6}{j+2} - \frac{2\,S_1(j+2)}{j+2}-1
\end{array}\nonumber
\end{eqnarray}
At LO we have three independent relations among elements of forward anomalous dimensions
matrix: Eq.~(\ref{DokSUSY}) together with Eq.~(\ref{SUSY2a}), Eq.~(\ref{SUSY2b})
and Eq.~(\ref{for-nonfora}), where at LO we have $\Delta^i_{j+1,j-1} = 0$.
Therefore, we may conclude, that in forward case in leading order of perturbation theory,
it is sufficient to know only one diagonal element of anomalous
dimensions matrix (\ref{MADMF1L}) or anomalous dimension of fermionic operator
${}^{\mathbf {fer}}\gamma_j$ to reconstruct the whole matrix.

At NLO the righthand side of Eq.~(\ref{for-nonfora}) is nonzero and as a consequence
matrices (\ref{MADMF2L}) and (\ref{MADMF2LP}) become triangular. So, to reconstruct
the forward anomalous dimensions matrix one needs to know additionally any other
element of forward anomalous dimensions matrix or non-diagonal anomalous dimension
of fermionic operator ${}^{\mathbf {fer}}\gamma_{j,j-1}$.

\section{Solution of NLO evolution equation}

Our primary goal is the construction of multiplicatively renormalized conformal operators up
to next-to-leading order in perturbation theory. The non-singlet case is simple and the corresponding
solution could be easily found in full analogy with $\phi^3$-theory~\cite{MRPhi36}.
So, here we will be interested only in singlet case with bosonic operator mixing.
In previous section we shown, that the twist-2 conformal operators form a chiral supermultiplet
of $N=1$ supersymmetry algebra and derived their transformation rules under supersymmetry.
It turns out, that in the singlet case, it is sufficient to find only the expression for multiplicatively
renormalized fermionic operator. The multiplicatively renormalized bosonic operators up to
next-to-leading order in perturbation theory could be obtained then via supersymmetry transformations
from the expression for multiplicatively renormalized fermionic operator.

As we already mentioned before there is a close relation between DA and conformal operators, we
are studying. To see it explicitly, let us consider the case of non-singlet bosonic operator.
The eigenvalues of the anomalous dimensions matrix $\hat\gamma$ of this operator are
$\gamma_j = \gamma_{j j}$. Therefore, the renormalization group analysis provides for
multiplicatively renormalized operators $\tilde{O}_{j l}$:
\be
\tilde{O}_{j l}(\mu^2) = \exp\left(\int_{\mu_0^2}^{\mu^2}\frac{dt}{t}\gamma_j\left(g (t)\right)\right)
\tilde{O}_{j l}(\mu_0^2).
\ee
Moreover, the operators $O_{j l}$ could be completely expressed through $\tilde{O}_{j l}$:
\be
O_{j l}(\mu^2) = \sum_{k=0}^{k} B_{jk}\bigl(g(\mu^2)\bigr)\tilde{O}_{k l}(\mu^2).
\ee
Then, the evolution of distribution amplitude is given by:
\be
\phi (x,Q^2) = \sum_{n = 0}^{\infty}\phi_n\!\!\left(x, \alpha (Q^2)\right)
\exp\left(\int_{Q_0^2}^{Q^2}\frac{dt}{t}\gamma_n\!\bigl(g(t)\bigr)\right)
\langle 0|O_{n n}(Q_0^2)|P\rangle^{\mathbf{red}},
\ee
where
\be
\phi_n\!\!\left(x,\alpha (Q^2)\right) =
\sum_{k = n}^{\infty}\frac{x\bar x}{{\cal N}_k}C_k^{3/2} (x-\bar x)B_{k n}\!\!\left(\alpha (Q^2)\right),
\ee
where ${\cal N}_k$ is some normalization factor (see below).
So, in this section we will first determine the eigenfunctions of the fermionic evolution
kernel. Then, it is straightforward to write down the multiplicatively renormalized fermionic
operators. Our treatment here is close to the work of Mikhailov and Radyushkin on $\phi^3$-theory
in 6-dimensional space-time~\cite{MRPhi36,MRKernel}.

In leading order of perturbation theory the evolution equation~(\ref{EvEq})
for the fermionic operator~(\ref{ConfOpFWZ1}) could be easily solved
employing separation of variables.
\begin{equation}\label{SolLO}
\left.\phi(x,\mu^2)\right|_{LO}=
\sum_{n=0}a_n\left(\ln\frac{\mu^2}{\Lambda^2}\right)^{-\left(^{\bf {fer}}\gamma^{(0)}_n/b_0\right)}
\!\!\!\bar x P^{(1,0)}_n(x-\bar x)\,,
\end{equation}
where $^{\bf {fer}}\gamma^{(0)}_n$ is the LO anomalous dimension of the
fermionic operator~(\ref{ConfOpFWZ1}) and $\bar x  P^{(1,0)}_n(x-\bar x)=\psi_n(x)$
is the solution of the eigenvalue equation
\begin{equation}
\int^1_0 v_0(x,y)\psi_n(y)d y=\lambda_n\psi_n(x)\,,\label{EigenValEq}
\end{equation}
where $\lambda_n=(-1)^n/(n+1)$ and $v_0$ is defined as
\begin{equation}
{\mbox {\bf V}}_{\!\!{\bf {fer}}}^{(0)}(x,y)=v_0(x,y)+\delta(x-y)
\end{equation}
The easiest way to find $\psi_n$'s is to note, that the combination $\bar y v_0 (x,y)\equiv \tilde{v}(x,y)$
is symmetric under the change $x\leftrightarrow y$: $\tilde{v}(x,y) = \tilde{v}(y,x)$. Another property
of $v_0 (x,y)$ is that its convolution with the polynomial of degree $N$ is the polynomial of degree $M$
with $M\leq N$. Thus, the functions $\psi_n(x)/\bar x$ should be the polynomials orthogonal to each other
on $[0,1]$ with measure $\bar x$.

Now, let us proceed with NLO evolution kernel. It is convenient to rewrite the resulting expression
Eq.~(\ref{EKFerNLO}) for NLO evolution kernel of the fermionic operators in the following form
\begin{eqnarray}
{\mbox {\bf V}}_{\!\!{\bf {fer}}}^{(1)} &=&
\theta (x > \bar y)\left\{\frac{1}{\bar y}\ln(x)
  - \frac{1}{\bar y}\ln\!\!\left(1-\frac{\bar x}{y} \right)\right\}
 + \theta (x < \bar y)\left\{\frac{1}{\bar y}\ln(\bar y - x)
 + \frac{6}{\bar y} \right\}
  \nonumber \\
&+&
\theta(x<y)\left\{ -\frac{\ln^2(y)}{\bar y} + \frac{2\ln(x)\ln(y)}{\bar y} \right\}
  + \theta (x>y)\left\{\frac{\ln^2(x)}{\bar y} \right\}\label{EKFerNLOrw}\\
&+&
\theta (x < \bar y)\left\{
   \frac{2}{\bar y}\ln\!\left(\frac{x}{\bar y}\right) \right\} +\delta(x-y)\nonumber
\end{eqnarray}
It is easy to verify, that the combination $\bar y{\mbox {\bf V}}_{\!\!{\bf {fer}}}^{(1)}$ is
symmetric with respect to transformation $x\leftrightarrow y$  if one takes into account
only the first line of Eq.~(\ref{EKFerNLOrw}). The second and third lines of the
expression above constitute non-symmetric piece and thus lead to the non-diagonal terms in the
Jacobi polynomial basis (excluding $\delta$-function, of course). To find a solution of
NLO evolution equation we will need the expression of NLO evolution kernel in the basis of LO
eigenfunctions. To simplify further notation it is convenient to introduce following definitions:
\begin{eqnarray}
\Psi^{(1,0)}_{\nu, n}(x) &=& \frac{w(1+\nu, \nu | x-\bar x)}{{\cal N}_{n}(1+\nu, \nu)}
(-1)^n P_{n}^{(1+\nu, \nu)}(x-\bar x)\label{PsiNu} \\
\Psi^{(1,0)}_{n}(x) &=& \left. \Psi^{(1,0)}_{\nu, n}(x)\right|_{\nu = 0},
\end{eqnarray}
where
${\cal N}_n(\alpha,\beta)=2^{\alpha+\beta+1}\frac{\Gamma(n+\alpha+1)\Gamma(n+\beta+1)}
{(2n+\alpha+\beta+1)\Gamma(n+1)\Gamma(n+\alpha+\beta+1)}$
is the normalization factor and
$w(\alpha,\beta,x)=(1-x)^\alpha(1+x)^\beta$ is weight function.
These are just the solutions of the eigenvalue equation~(\ref{EigenValEq})
and its generalization:
\begin{equation}
\int_0^1 dy V_{\nu}^{(0)}(x,y)
\Psi^{(1,0)}_{\nu, n}(y)=-
\frac{(-1)^{n+1}}{n+1+\nu}
\Psi^{(1,0)}_{\nu, n}(y)\,,
\end{equation}
where
\begin{equation}
V_{\nu}^{(0)}(x,y) = \theta(x<\bar y)\frac{1}{\bar y}\left(\frac{x}{\bar y}\right)^{\nu}
\end{equation}
To write down the evolution kernel in the basis of LO eigenfunctions, it is convenient
to express the NLO kernel as the convolution of LO evolution kernels and their
derivatives:
\begin{eqnarray}
v_0(x,y) &=& \theta(x<\bar y)\frac{1}{\bar y}\\
\dot{v}_0(x,y) &=& \left.\left(\frac{d}{d\nu} ~V_{\nu}^{(0)}(x,y)\right)\right|_{\nu = 0}
= \theta(x<\bar y)\frac{1}{\bar y}\ln\left(\frac{x}{\bar y}\right)\label{v0dot}
\end{eqnarray}
Then it is an easy exercise to derive, that:
\begin{equation}\label{EKFerNLOV1}
V^{(1)}(x,y) = -2\dot{v}_0(x,z)\otimes v_0(z,y) +
6v_0(x,y) + 2\dot{v}_0(x,y) + u(x,y)\,,
\end{equation}
where $\otimes$ denotes the convolution and
\begin{eqnarray}
u(x,y) &=&
\theta (x > \bar y)\left\{\frac{1}{\bar y}\ln(x)
- \frac{1}{\bar y}\ln\!\!\left(1-\frac{\bar x}{y} \right)\right\}
+ \theta (x < \bar y)\left\{\frac{1}{\bar y}\ln(\bar y - x)\right\}
\end{eqnarray}
It is not hard to show that $u(x,y)$ is diagonal in the basis of LO eigenfunctions:
\begin{eqnarray}
\int_0^1 dy\; u(x,y)\Psi^{(1,0)}_{n}(y) = (-1)^{n+1}\left\{-\frac{1}{(n+1)^2} + \frac{2 S_1(n+1)}{n+1}
\right\}\Psi^{(1,0)}_n(x).
\end{eqnarray}
Now, what is left is to specify the action of dot kernels on LO eigenfunctions. As a starting point
we take the following eigenvalue equation
\begin{equation}
V_{\nu}^{(0)}(x,y)\otimes\Psi^{(1,0)}_{\nu, n}(y) =
-\frac{(-1)^{n+1}}{n+1+\nu}\Psi^{(1,0)}_{\nu, n}(x)
\end{equation}
Differentiating this equation with respect to $\nu$ and putting it afterwards to zero
we get
\begin{eqnarray}
&& \dot{v}_0(x,y)\otimes\Psi^{(1,0)}_{n}(y)
+ v_0(x,y)\otimes \left.\frac{d}{d\nu}\left(\Psi^{(1,0)}_{\nu, n}(y) \right)\right|_{\nu = 0}
= \nonumber \\ && \frac{(-1)^{n+1}}{(n+1)^2}\Psi^{(1,0)}_{n}(x) -
\frac{(-1)^{n+1}}{n+1}\left.\frac{d}{d\nu}\left(\Psi^{(1,0)}_{\nu,n}(x)\right)\right|_{\nu = 0}
\label{DotEigValEq}
\end{eqnarray}
So, to find an action of dot kernels on LO eigenfunctions we need to
know an expression for the $\nu$ derivative of  $\Psi^{(1,0)}_{\nu, n}$. The latter is given by
\begin{equation}\label{ExpDot}
\left.\frac{d}{d\nu}\left(\Psi^{(1,0)}_{\nu, n}(x) \right)\right|_{\nu = 0} =
2 \sum_{k>n}^{\infty} \hat d_{kn} \Psi^{(1,0)}_k(x) + \mbox{diagonal part}\,,
\end{equation}
where
\begin{eqnarray}\label{hatdkn}
\hat d_{jk} = \frac{1}{j-k}\frac{(j+1)+(-1)^{j-k}(k+1)}{j+k+2}\frac{k+1}{j+1}
\end{eqnarray}
We failed to find diagonal part in the equation above, however it is quite easy to
derive diagonal part directly for $\dot{v}_0(x,y)\otimes\Psi^{(1,0)}_{n}(y)$. In this way, we get
from Eq.~(\ref{DotEigValEq})
\begin{equation}
\dot{v}_0(x,y)\otimes\Psi^{(1,0)}_{n}(y) = \frac{(-1)^{n+1}}{(n+1)^2}\Psi^{(1,0)}_{n}(x)
- \sum_{k > n}^{\infty} d_{kn}(\gamma^{(0)}_k-\gamma^{(0)}_n)\Psi^{(1,0)}_{k}(x)
\end{equation}
We can summarized the actions of building blocks of NLO evolution kernel on LO eigenfunctions
as follows
\begin{eqnarray}
v_0(x,y)\otimes\Psi^{(1,0)}_{n}(y) &=& -\frac{(-1)^{n+1}}{n+1}\Psi^{(1,0)}_{n}(x) \\
\dot{v}_0(x,y)\otimes\Psi^{(1,0)}_{n}(y) &=& \frac{(-1)^{n+1}}{(n+1)^2}\Psi^{(1,0)}_{n}(x)
- \sum_{k > n}^{\infty} d_{kn}(\gamma^{(0)}_k-\gamma^{(0)}_n)\Psi^{(1,0)}_{n}(x) \\
u(x,y)\otimes\Psi^{(1,0)}_{n}(y) &=&
(-1)^{n+1}\left\{-\frac{1}{(n+1)^2}+\frac{2 S_1(n+1)}{n+1} \right\}\Psi^{(1,0)}_{n}(x)
\end{eqnarray}
With all these formula at hand it is easy to derive that
\begin{eqnarray}
-2\dot{v}_0(x,z)\otimes v_0(z,y)\otimes\Psi^{(1,0)}_{n}(y) &=& \frac{2}{(n+1)^3}\Psi^{(1,0)}_{n}(x)\\
&+& \sum_{k>n}^{\infty}
d_{kn}\left(\gamma^{(0)}_k - \gamma^{(0)}_n\right)
\left(-1-\gamma^{(0)}_n\right)\Psi^{(1,0)}_{k}(x)\nonumber
\end{eqnarray}
Note, that substituting the expression for NLO evolution kernel in the basis of LO eigenfunction
into Eq.~(\ref{EKADRelFer}) we can easily determine analytical expression for
non-diagonal part of anomalous dimensions matrix of fermionic operator.

Now we are ready to write down the solution of ER-BL evolution equation
for fermionic operator (\ref{ConfOpFWZ1}) in next-to-leading order.
Note, first, that if the next-to-leading order kernel $V^{(1)}(x,y)$ is diagonal in the Jacobi polynomials basis,
then the solution of evolution equation (\ref{EvEq}) would be
\begin{equation}\label{SolDiag2}
\phi^{\mathrm{(diag)}}_n\left(x,\mu^2\right)=a_n\left(\mu^2_0\right)
exp\left(-\int^{\mu^2}_{\mu^2_0}{}^{\bf{fer}}\gamma_n(g(t))\frac{dt}{t}\right)\bar xP^{(1,0)}_n(x-\bar x),
\end{equation}
where ${}^{\bf{fer}}\gamma_n(g)=\left(\frac{\alpha}{4\pi}\right){}^{\bf{fer}}\gamma^{(0)}_n+
\left(\frac{\alpha}{4\pi}\right)^2{}^{\bf{fer}}\gamma^{(1)}_n$ with ${}^{\bf{fer}}\gamma^{(0)}_n$ and
${}^{\bf{fer}}\gamma^{(1)}_n$ given by Eqs.~(\ref{AD1LFer}) and (\ref{AD2LFer}) correspondingly.
In the case when $V^{(1)}(x,y)$ contains the non-diagonal terms, the solution of the
evolution equation will differ from $\phi^{\mathrm{(diag)}}_n$ by ${\cal O}(\alpha)$ terms.
Therefore, we will look for the solution in the following form
\begin{equation}\label{NLOpsi}
\phi_n=\left(1+\frac{\alpha}{4\pi} W\right)\otimes \phi^{\mathrm{(diag)}}_n,
\end{equation}
Consider first a more simple case $b_0 = 0$. Substituting (\ref{NLOpsi}) into the evolution equation and using the
explicit form of the total kernel (\ref{EKFerNLOV1}) one can find, that in this case
$W=W(x,y)$ should satisfy the following equation
\begin{equation}\label{eqforW}
\left[v_0,W\right]_-+\dot v_0\otimes\left(\gamma_\psi-v_0\right)=0.
\end{equation}
This equation could be easily solved noticing, that it is similar to eigenvalue equation
for the $V^{(0)}_\nu$ kernel written up to ${\cal O}(\nu)$ terms
\begin{equation}
\left[v_0,\omega\right]_-+\dot v_0=0,
\end{equation}
where $\omega$ stands for the generator of $\nu$-shifts:
\begin{equation}
\Psi^{(1,0)}_{\nu,n}\equiv\Psi^{(1,0)}_{n}+\left.\nu\frac{d}{d\nu}\Psi^{(1,0)}_{\nu,n}\right|_{\nu=0}
+{\cal O}(\nu^2)=(1+\nu\;\omega)\otimes\Psi^{(1,0)}_{n}+{\cal O}(\nu^2)
\end{equation}
As a result, we have the following expression for $W$:
\begin{equation}
W=-\omega\otimes\left(v_0-\gamma_\psi\right)
\end{equation}
and as a consequence we have
\begin{eqnarray}
\phi_n&=&\left(1+\frac{\alpha}{4\pi}\, W\right)\otimes\Psi^{(1,0)}_n=
\Bigl(1-\frac{\alpha}{4\pi}\,\bigl(\lambda_n-\gamma_\psi\bigr)\omega\Bigr)\otimes\Psi^{(1,0)}_n\nonumber\\
&=&\left.\Psi^{(1,0)}_{\nu,n}(x)\right|_{\nu=\frac{\alpha}{4\pi}\,\gamma^{(0)}_n}
\Bigl(1+{\cal O}\!\left(\alpha^2\right)\Bigr)
\end{eqnarray}
Then, the multiplicatively renormalized operators in this ($b_0=0$) case could be written through the
coefficients $\hat d_{kn}$ (\ref{hatdkn}) of the expansion (\ref{ExpDot}):
\begin{equation}
O_n={\cal O}_n+\frac{\alpha}{4\pi}\sum_{n>k}{\hat d}_{kn}\gamma^{(0)}_k{\cal O}_n
\end{equation}
In the $b_0\neq 0$ case the equation for the rotation kernel $W$ has even more
complicated structure than Eq.~(\ref{eqforW})
\begin{equation}\label{eqforWb0}
b_0W+\left[v_0,W\right]_-+\dot v_{0}\otimes\left(b_0+\gamma_\psi-v_0\right)=0.
\end{equation}
The effective method to solve this equation is to use matrix representation for the
kernels in the Jacobi polynomial basis. It then follows, that $W\otimes \Psi^{(1,0)}_n$ has
the following form:
\begin{equation}
W\otimes \Psi^{(1,0)}_n=\left(b_0+\gamma_\Phi-\lambda_n\right)\left(\lambda_n-v_0-b_0\right)^{-1}
\otimes\dot v_0\otimes\Psi^{(1,0)}_n
\end{equation}
Now using the expression for the dot kernel in the Jacobi polynomial basis derived previously, we
get
\begin{equation}
W\otimes\Psi^{(1,0)}_n=\left(b_0+{}^{\bf{fer}}\gamma^{(0)}_n\right)
\left(\bar P_n \dot{\Psi}^{(1,0)}_n +b_0\sum^\infty_{k>n}
\frac{\hat d_{kn}}{{}^{\bf{fer}}\gamma^{(0)}_k-{}^{\bf{fer}}\gamma^{(0)}_n-b_0}
\Psi^{(1,0)}_k\right),
\end{equation}
where  $\bar P_n$ is the projection operator $\bar P_n \Psi^{(1,0)}_n=(1-\delta_{nk})\Psi^{(1,0)}_n$
subtracting the diagonal part. The multiplicatively renormalized operators in this case are given by
\be
O_n = {\cal O}_n + \frac{\alpha}{4\pi}(b_0 + {}^{\bf{fer}}\gamma^{(0)}_n)
\sum_{n > k}\hat{d}_{k n}\frac{{}^{\bf{fer}}\gamma^{(0)}_k - {}^{\bf{fer}}\gamma^{(0)}_n}
{{}^{\bf{fer}}\gamma^{(0)}_k - {}^{\bf{fer}}\gamma^{(0)}_n - b_0}{\cal O}_k.
\ee

\section{Conclusion}

It follows from our analysis, that to find multiplicatively renormalized twist-2 conformal operators
in Wess-Zumino model it is sufficient to find multiplicatively renormalized operators only for
one member of operator supermultiplet. It is convenient in this case to find a solution for fermionic
operator. For the latter the problem of finding multiplicatively renormalized operators is much simpler
compared to singlet bosonic operators, where we have operator mixing. All other multiplicatively renormalized
operators could be then obtained from fermionic operator with the use of supersymmetry transformations.
Moreover, we found, that the knowledge of fermionic diagonal and non-diagonal anomalous dimensions matrices
allows us completely reconstruct the forward anomalous dimensions matrix in singlet case.
Also, the analysis was performed in a simplified supersymmetric Wess-Zumino model, we think that most
of our findings could be straightforwardly applied to field theories with more supersymmetries.

\vspace{1cm} 
{\large \bf Acknowledgments} 

\vspace*{0.5cm}
\noindent
We are grateful to A.V.~Kotikov, L.N.~Lipatov and S.V.~Mikhailov for numerous discussions on the subject of this
paper. The work of A.O. was supported by the National Science Foundation under grant PHY-0244853 and by
the US Department of Energy under grant DE-FG02-96ER41005. The work of V.V. 
is supported by grants INTAS 00-366 and RSG SS-1124.2003.2.

\section*{Appendix A}

Here we give the diagram by diagram results. The whole
calculation was automized with the help of Feynman diagram
analyzer DIANA~\cite{DIANA}, calling QGRAF~\cite{QGRAF} for the Feynman
diagram generation, and computer algebra system FORM~\cite{FORM}.

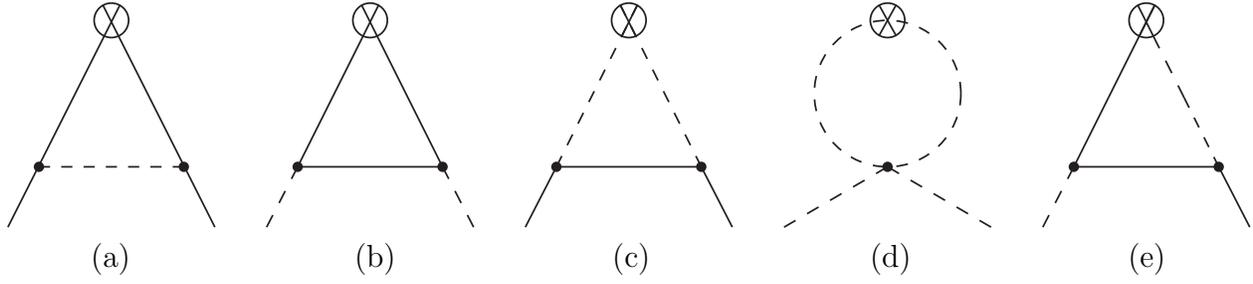
\begin{figure}[ht]
\vspace{-1.5cm}
\begin{center}
\begin{picture}(680,120)(0,0)
\SetScale{0.65}
\SetWidth{1}
\Line(40,-20)(105,109)
\Line(160,-20)(95,109)
\CArc(100,100)(10,0,360)
\DashLine(142,15)(58,15){7}
\Vertex(142,15){3}
\Vertex(58,15){3}
\Text(65,-25)[c]{(a)}

\Line(208,15)(255,109)
\DashLine(190,-20)(208,15){7}
\Line(292,15)(245,109)
\DashLine(310,-20)(292,15){7}
\CArc(250,100)(10,0,360)
\Line(292,15)(208,15)
\Vertex(292,15){3}
\Vertex(208,15){3}
\Text(165,-25)[c]{(b)}

\Line(405,109)(396,91)
\Line(404,91)(395,109)
\DashLine(358,15)(405,109){7}
\Line(340,-20)(358,15)
\DashLine(442,15)(395,109){7}
\Line(460,-20)(442,15)
\CArc(400,100)(10,0,360)
\Line(442,15)(358,15)
\Vertex(442,15){3}
\Vertex(358,15){3}
\Text(262,-25)[c]{(c)}

\Line(545,91)(555,109)
\DashLine(490,-20)(550,15){7}
\Line(555,91)(545,109)
\DashLine(610,-20)(550,15){7}
\DashCArc(550,57.5)(42.5,0,360){7}
\CArc(550,100)(10,0,360)
\Vertex(550,15){3}
\Text(360,-25)[c]{(d)}

\Line(705,109)(696,91)
\Line(704,91)(695,109)
\DashLine(640,-20)(658,15){7}
\Line(658,15)(678,55)
\Line(678,55)(705,109)
\Line(760,-20)(742,15)
\DashLine(742,15)(722,55){7}
\DashLine(722,55)(704,91){7}
\CArc(700,100)(10,0,360)
\Line(742,15)(658,15)
\Vertex(742,15){3}
\Vertex(658,15){3}
\Text(457,-25)[c]{(e)}
\end{picture}
\end{center}
\vspace{10mm}
\caption{One-loop diagrams}
\label{OneLoop}
\end{figure}

The results for each diagram include contributions of this diagram to
forward anomalous dimension $\gamma_m$, to forward
evolution kernel ${\mbox {\bf P}}(z)$ and to non-forward evolution kernel ${\mbox {\bf V}}(x,y)$.
Here and below, the contributions of diagrams marked by star '$*$' change the overall
sign in polarized case. The tadpole diagrams (similar to~\ref{OneLoop}.d) give zero contribution in
forward case and are proportional to constant in non-forward case. So, in what follows we do
not write explicit expressions for contributions of these diagrams. Also, graphs containing external
self-energies, giving $\delta$-function contribution to evolution kernels, were taken into
account in calculation, but not shown explicitly here.

\newpage

\renewcommand{\arraystretch}{0}
\begin{center}
\begin{equation*}
\begin{array}{||c||c||}
\hline
&\\[0.5mm]
\hline
&\\[2mm]
\qquad\qquad&\quad\displaystyle{\frac{2}{m+1}-\frac{2}{m+2}}\\
&\\[2mm]
\cline{2-2}
&\\[2mm]
1.a\;{}^*&\quad 2 \bar{z}\\
&\\[2mm]
\cline{2-2}
&\\[2mm]
&\quad\displaystyle{\theta ( x < y ) 2\frac{x}{y}
+ \left(\begin{array}{cc}
    x \leftrightarrow \bar{x},&
    y \leftrightarrow \bar{y}
  \end{array}\right)}\\
&\\[2mm]
\hline
&\\[0.5mm]
\hline
&\\[2mm]
&\quad\displaystyle{\frac{2}{m+1}\frac{m+1}{2}}\\
&\\[2mm]
\cline{2-2}
&\\[2mm]
1.b\;{}^*&\quad 2\\[2mm]
\cline{2-2}
&\\[2mm]
&\quad\displaystyle{\theta ( x < y ) 2 x
-  \left(\begin{array}{cc}
    x \leftrightarrow \bar{x},&
    y \leftrightarrow \bar{y}
  \end{array}\right)}\\
&\\[2mm]
\hline
&\\[0.5mm]
\hline
&\\[2mm]
&\quad\displaystyle{\frac{2}{m+2}\frac{2}{m+1}}\\
&\\[2mm]
\cline{2-2}
&\\[2mm]
1.c\;{}^*&\quad 2 z\\
&\\[2mm]
\cline{2-2}
&\\[2mm]
&\quad\displaystyle{\theta ( x < y ) \frac{2}{y}
-  \left(\begin{array}{cc}
    x \leftrightarrow \bar{x},&
    y \leftrightarrow \bar{y}
  \end{array}\right)}\\
&\\[2mm]
\hline
&\\[0.5mm]
\hline
&\\[2mm]
&\quad 0\\
&\\[2mm]
\cline{2-2}
&\\[2mm]
1.d&\quad 0\\
&\\[2mm]
\cline{2-2}
&\\[2mm]
&\quad \displaystyle{\theta ( x < y )
 +  \left(\begin{array}{cc}
    x \leftrightarrow \bar{x},&
    y \leftrightarrow \bar{y}
  \end{array}\right)}\\
&\\[2mm]
\hline
&\\[0.5mm]
\hline
&\\[2mm]
&\quad\displaystyle{(-1)^{m+1}\frac{2}{m+1}}\\
&\\[2mm]
\cline{2-2}
&\\[2mm]
1.e\;{}^*&\quad 2 z \\
&\\[2mm]
\cline{2-2}
&\\[2mm]
&\quad \displaystyle{\theta ( x < \bar y ) \frac{2}{\bar y}}\\
&\\[2mm]
\hline
&\\[0.5mm]
\hline
\end{array}
\end{equation*}
\end{center}

\newpage

\renewcommand{\arraystretch}{0}
\begin{center}
\begin{equation*}
\begin{array}{||c||c||}
\hline
&\\[0.5mm]
\hline
&\\[3mm]
\qquad\qquad&\quad\displaystyle{-\frac{2}{(m+1)^3}+\frac{4}{(m+1)^2}-\frac{6}{m+1}
-\frac{2}{(m+2)^3}+\frac{6}{m+2}}\\
&\\[3mm]
\cline{2-2}
&\\[3mm]
2.a&\quad -6 \bar{z} - 4 \ln ( z ) - ( 1 + z ) \ln^2 ( z )\\
&\\[3mm]
\cline{2-2}
&\\[3mm]
&\quad\displaystyle{\theta ( x < y ) \left\{4 \frac{x}{y} + 2\frac{\bar{x}}{y} \ln ( \bar{x} )
+2 \frac{x}{\bar{y}} \ln ( y ) +  \frac{\bar{x}}{y} \ln^2 ( \bar{x} )
+2 \frac{x}{\bar{y}} \ln ( x ) \ln ( y )\right.}\\[5mm]
&\displaystyle{\left.   - \frac{x}{\bar{y}} \ln^2 ( y )\right\}
+ \left(\begin{array}{c}
    x \leftrightarrow \bar{x}\\[3mm]
    y \leftrightarrow \bar{y}
  \end{array}\right)}\\
&\\[3mm]
\hline
&\\[0.5mm]
\hline
&\\[3mm]
&\quad\displaystyle{-\frac{2}{(m+1)^2}+\frac{6}{m+1}-\frac{2}{(m+2)^2}-\frac{6}{m+2}}\\
&\\[3mm]
\cline{2-2}
&\\[3mm]
2.b\;{}^*&\quad6\bar z+2(1+z)\ln(z)\\[3mm]
\cline{2-2}
&\\[3mm]
&\quad\displaystyle{\theta ( x < y ) \left\{-4 \frac{x}{y} -2 \frac{\bar{x}}{y} \ln ( \bar{x} )
-2 \frac{x}{\bar{y}} \ln ( y )\right\}+
  \left(\begin{array}{c}
    x \leftrightarrow \bar{x}\\[3mm]
    y \leftrightarrow \bar{y}
  \end{array}\right)}\\
&\\[3mm]
\hline
&\\[0.5mm]
\hline
&\\[3mm]
&\quad\displaystyle{ \frac{4}{m+1}-\frac{4}{m+2}-2\frac{S_1(m+1)}{m+1}+2\frac{S_1(m+2)}{m+2}}\\
&\\[3mm]
\cline{2-2}
&\\[3mm]
2.c&\quad 2\bar z\left(2+\ln(\bar z)\right)\\
&\\[3mm]
\cline{2-2}
&\\[3mm]
&\quad\displaystyle{\theta ( x < y ) \left\{-4\frac{x}{y} - \frac{\bar{x}}{\bar{y}} \ln ( \bar{x} )
- \frac{x}{y} \ln ( x )+ \frac{1}{\bar{y}}  \left( 1 - \frac{x}{y} \right) \ln \left( 1 - \frac{x}{y} \right)\right\}
+  \left(\begin{array}{c}
    x \leftrightarrow \bar{x}\\[3mm]
    y \leftrightarrow \bar{y}
  \end{array}\right)}\\
&\\[3mm]
\hline
&\\[0.5mm]
\hline
&\\[3mm]
&\quad\displaystyle{-\frac{1}{(m+1)^2}+\frac{3}{m+1}+\frac{1}{(m+2)^2}-\frac{3}{m+2}}\\
&\\[3mm]
\cline{2-2}
&\\[3mm]
2.d\;{}^*&\quad \bar z \left(3+\ln(z)\right)\\
&\\[3mm]
\cline{2-2}
&\\[3mm]
&\quad \displaystyle{\theta ( x < y )\; \frac{x}{y}\left\{- 2  -
  \ln\!\left ( \frac{x}{y} \right) \right\} +
  \left(\begin{array}{c}
    x \leftrightarrow \bar{x}\\[3mm]
    y \leftrightarrow \bar{y}
  \end{array}\right)}\\
&\\[3mm]
\hline
&\\[0.5mm]
\hline
\end{array}
\end{equation*}
\end{center}

\newpage

\vspace*{20mm}
\begin{equation*}
\begin{array}{||c||c||}
\hline
&\\[0.5mm]
\hline
&\\[3mm]
\qquad\qquad&\quad
\displaystyle{\left(-\frac{2}{(m +1)^3} + \frac{2}{( m + 1 )^2}- \frac{2}{( m + 2 )^2}\right)\frac{m+1}{2}} \\
&\\[3mm]
\cline{2-2}
&\\[3mm]
3.a&\quad \left(-2\bar z-\ln(z)\right)\ln ( z ) \\
&\\[3mm]
\cline{2-2}
&\\[3mm]
&\quad \displaystyle{ \theta ( x < y ) \Bigl\{ 2\bar{x} \ln ( \bar{x} ) +
  \bar{x} \ln^2 ( \bar{x} ) -2 x \ln ( x ) \ln ( y ) +
  x \ln^2 ( y )}\\[5mm]
&\quad\qquad\qquad\quad\displaystyle{ -2 x \ln ( x ) \Bigr\} - \left(\begin{array}{c}
    x \leftrightarrow \bar{x}\\[3mm]
    y \leftrightarrow \bar{y}
  \end{array}\right)}\\
&\\[3mm]
\hline
&\\[0.5mm]
\hline
&\\[3mm]
&\quad \displaystyle{ \left(\frac{4}{m+1} - \frac{2S_1 ( m + 1 )}{m+1}\right)\frac{m+1}{2}} \\
&\\[3mm]
\cline{2-2}
&\\[3mm]
3.b\;{}^*&\quad \displaystyle{ 2\left(2+\ln(\bar z)\right)}\\
&\\[3mm]
\cline{2-2}
&\\[3mm]
&\quad\displaystyle{{\bar x}\Bigl\{2+\ln({\bar x})\Bigr\}
  -\theta(y>x)\left\{2+\ln\!\left(1-\frac{x}{y}\right)\right\}}- \left(\begin{array}{c}
    x \leftrightarrow \bar{x}\\[3mm]
    y \leftrightarrow \bar{y}
  \end{array}\right)\\
&\\[3mm]
\hline
&\\[0.5mm]
\hline
&\\[3mm]
&\quad \displaystyle{ \left(-\frac{1}{(m+1)^2}+\frac{3}{m+1} - \frac{1}{m + 2}\right)\frac{m+1}{2}} \\
&\\[3mm]
\cline{2-2}
&\\[3mm]
3.c\;{}^*&\quad \displaystyle{3-z+\ln(z)} \\
&\\[3mm]
\cline{2-2}
&\\[3mm]
&\quad \displaystyle{\theta(y>x)\;x\!\left\{-2-\ln\!\left(\frac{x}{y}\right)\right \}
- \left(\begin{array}{c}
    x \leftrightarrow \bar{x}\\[3mm]
    y \leftrightarrow \bar{y}
  \end{array}\right)}\\
&\\[3mm]
\hline
&\\[0.5mm]
\hline
\end{array}
\end{equation*}

\newpage

\begin{center}
\begin{equation*}
\begin{array}{||c||c||}
\hline
&\\[0.5mm]
\hline
&\\[3mm]
\qquad\qquad&\quad
\displaystyle{\left(-\frac{2}{(m +1)^2} + \frac{2}{( m + 2 )^3} + \frac{2}{( m + 2 )^2}\right)\frac{2}{m+1}}\\
&\\[3mm]
\cline{2-2}
&\\[3mm]
4.a&\quad  z\ln^2(z)+2 \bar{z} \ln ( z )\\
&\\[3mm]
\cline{2-2}
&\\[3mm]
& \quad  \displaystyle{\theta ( x < y ) \left\{  \frac{1}{y}
  \ln^2 ( \bar{x} ) - \frac{2}{\bar{y}} \ln ( x ) \ln ( y ) +
  \frac{1}{\bar{y}} \ln^2 ( y ) \right\} - \left(\begin{array}{c}
    x \leftrightarrow \bar{x}\\[3mm]
    y \leftrightarrow \bar{y}
  \end{array}\right)}\\
&\\[3mm]
\hline
&\\[0.5mm]
\hline
&\\[3mm]
& \displaystyle{\left( \frac{4}{m + 2} - \frac{2 S_1 ( m + 2 )}{m + 2}\right)\frac{2}{m+1}}\\
&\\[3mm]
\cline{2-2}
&\\[3mm]
4.b\;{}^*& 2z\left(2+\ln ( \bar{z} )\right) \\
&\\[3mm]
\cline{2-2}
&\\[3mm]
&\qquad\displaystyle{-\frac{1}{\bar y}\Bigl\{2+\ln({\bar x})\Bigr\}
  +\theta(y>x)\frac{1}{y\bar y}\left\{2+\ln\!\left(1-\frac{x}{y}\right)\right\}}- \left(\begin{array}{c}
    x \leftrightarrow \bar{x}\\[3mm]
    y \leftrightarrow \bar{y}
  \end{array}\right)\\
&\\[3mm]
\hline
&\\[0.5mm]
\hline
&\\[3mm]
& \displaystyle{\left(-\frac{1}{m + 1} + \frac{1}{( m + 2 )^2} - \frac{3}{m + 2}\right)\frac{2}{m+2}} \\
&\\[3mm]
\cline{2-2}
&\\[3mm]
4.c\;{}^*& -1+ 3 z+z \ln ( z ) \\
&\\[3mm]
\cline{2-2}
&\\[3mm]
&\qquad \displaystyle{\theta(y>x)\;\frac{1}{y}\!\left\{2+\ln\!\left(\frac{x}{y}\right)\right\}
    - \left(\begin{array}{c}
    x \leftrightarrow \bar{x}\\[3mm]
    y \leftrightarrow \bar{y}
  \end{array}\right)
}\\
&\\[3mm]
\hline
&\\[0.5mm]
\hline
\end{array}
\end{equation*}
\end{center}
\newpage

\begin{center}
\begin{equation*}
\begin{array}{||c||c||}
\hline
&\\[0.5mm]
\hline
&\\[3mm]
\qquad\qquad& \displaystyle{-\frac{2}{m + 2}+\frac{2}{( m + 2 )^2} + \frac{2}{m + 1} - \frac{2}{( m +  1 )^2}} \\
&\\[3mm]
\cline{2-2}
&\\[3mm]
5.a& 2\bar{z} \left( 1 + \ln ( z )\right) \\
&\\[3mm]
\cline{2-2}
&\\[3mm]
&  \quad   \displaystyle{\theta ( x < y ) \left\{ -4 - 2\ln\! \left( \frac{x}{y} \right)  +
  \ln^2 ( \bar{x} ) + 2 \ln ( x ) \ln ( y ) - \ln^2 (y) \right\}
   + \left(\begin{array}{c}
    x \leftrightarrow \bar{x}\\[3mm]
    y \leftrightarrow \bar{y}
  \end{array}\right)}\\
&\\[3mm]
\hline
&\\[0.5mm]
\hline
&\\[3mm]
& \displaystyle{\frac{3}{m + 2} - \frac{3}{m + 1}}\\
&\\[3mm]
\cline{2-2}
&\\[3mm]
5.b& -3\bar{z} \\
&\\[3mm]
\cline{2-2}
&\\[3mm]
&  \quad \displaystyle{\theta ( x < y ) \left\{ 6 +3 \ln\! \left( \frac{x}{y} \right) \right\}
    + \left(\begin{array}{c}
    x \leftrightarrow \bar{x}\\[3mm]
    y \leftrightarrow \bar{y}
  \end{array}\right)}\\
&\\[3mm]
\hline
&\\[0.5mm]
\hline
\end{array}
\end{equation*}
\end{center}

\begin{center}
\begin{equation*}
\begin{array}{||c||c||}
\hline
&\\[0.5mm]
\hline
&\\[3mm]
\qquad\qquad& \displaystyle{\frac{2}{(m + 1)^3}-\frac{2}{(m + 1)^2}} \\
&\\[3mm]
\cline{2-2}
&\\[3mm]
6.a& (2+\ln(z))\ln(z) \\
&\\[3mm]
\cline{2-2}
&\\[3mm]
&  \quad \displaystyle{\theta ( x < y )
   \left\{2\frac{\ln(x)\ln(y)}{\bar y}+
   2\frac{\ln(y)}{\bar y}-
   \frac{\ln^2(y)}{\bar y}
   \right\} +
   \theta ( y < x )
   \left\{2\frac{\ln(x)}{\bar y}+
   \frac{\ln^2(x)}{\bar y}
   \right\}} \\
&\\[3mm]
\hline
&\\[0.5mm]
\hline
&\\[3mm]
& \displaystyle{(-1)^{m+1}\left(-\frac{2}{m + 1} + \frac{S_1(m+1)}{m + 1}\right)} \\
&\\[3mm]
\cline{2-2}
&\\[3mm]
6.b& -2-\ln(\bar z) \\
&\\[3mm]
\cline{2-2}
&\\[3mm]
&  \quad \displaystyle{\theta ( x < \bar y )
   \frac{2}{\bar y} -
   \theta ( x > \bar y )
   \frac{1}{\bar y}\ln\left(1-\frac{\bar x}{y}\right) +
   \frac{1}{\bar y}\ln(x)}\\
&\\[3mm]
\hline
&\\[0.5mm]
\hline
&\\[3mm]
& \displaystyle{(-1)^{m+1}\left(\frac{1}{(m + 1)^2} - \frac{2}{m + 1} \right)}\\
&\\[3mm]
\cline{2-2}
&\\[3mm]
6.c& -2-\ln(z) \\
&\\[3mm]
\cline{2-2}
&\\[3mm]
&  \quad \displaystyle{\theta ( x < \bar y )\left\{
   \frac{2}{\bar y} +
   \frac{1}{\bar y}\ln\left(\frac{x}{\bar y}\right)\right\} } \\
&\\[3mm]
\hline
&\\[0.5mm]
\hline
&\\[3mm]
& \displaystyle{(-1)^{m+1}\left(-\frac{2}{m + 1} + \frac{S_1(m+1)}{m + 1}\right)} \\
&\\[3mm]
\cline{2-2}
&\\[3mm]
6.d& -2-\ln(\bar z) \\
&\\[3mm]
\cline{2-2}
&\\[3mm]
&  \quad \displaystyle{\theta ( x < \bar y )\left\{
   \frac{2}{\bar y} +
   \frac{1}{\bar y}\ln\left(1-\frac{x}{\bar y}\right)\right\}}  \\
&\\[3mm]
\hline
&\\[0.5mm]
\hline
&\\[3mm]
& \displaystyle{\frac{2}{(m + 1)^2}} \\
&\\[3mm]
\cline{2-2}
&\\[3mm]
6.e& -2\ln(z) \\
&\\[3mm]
\cline{2-2}
&\\[3mm]
&  \quad \displaystyle{-\theta ( x < y )
   \frac{2}{\bar y}\ln(y) -
   \theta ( y < x )
   \frac{2}{\bar y}\ln(x)    }\\
&\\[3mm]
\hline
&\\[0.5mm]
\hline
\end{array}
\end{equation*}
\end{center}

\section*{Appendix B}

Here we present details on the evaluation of momentum integrals present in
the calculation of evolution kernels of ER-BL equations. Our consideration
follows the original work of Mikhailov and Radyushkin~\cite{MRQCD}\footnote{It
should be noted that expressions for $J(z)$ and $S(z)$ functions (to be introduced
below) have misprints in \cite{MRQCD}. We thank Sergei Mikhailov for fruitful discussions on this
subject and performed crosschecks.}

The main complication of the integrals involved compared to usual propagator-type integrals (for which
a lot of powerful technics were developed over the past years) is the presence
of $\delta$-function related to effective vertex. To derive all the formula we
used $\alpha$-presentation for momentum integrals as was originally proposed
in \cite{MRQCD}. For propagators and $\delta$-function insertion we have:
\be
\frac{1}{(k^2+i\ep)^{\nu}} &=& \frac{(-i)^{\nu}}{\Gamma (\nu)}
\int_0^{\infty}d\alpha\alpha^{\nu -1}\cdot\exp \Bigl\{i\alpha \Bigl(k^2 + i\ep\Bigr)\Bigr\} \\
\delta\left(x-\frac{k\cdot n}{P\cdot n} \right) &=&
\frac{1}{2\pi}\int_{-\infty}^{\infty} d\alpha
\exp\left\{i\alpha \left(x-\frac{k\cdot n}{P\cdot n}\right)  \right\}
\ee
In what follows we used the same integration measure convention as used
by \cite{MRQCD}, that is we use $\overline{\mbox{MS}}$ prescription. Technically, it
could be accomplished by multiplying each integration $d^d k$ by additional
factor $M(\ep) = (4\pi)^{-\ep}\Gamma (1-\ep)$. At one-loop level all integrals
obtained after projecting out Dirac structure of the conformal operator under
consideration  and expanding scalar products of momenta in the numerator over
the denominator factors could be reduced to the following two most general integrals
\be
&& \int\frac{d^d k}{(2\pi)^d}\frac{\delta\left(1-\frac{n\cdot k}{n\cdot P} \right)}
{[(k - a P)^2 - m_1^2]^{n_1}[(k - b P)^2 - m_2^2]^{n_2}} = \nonumber \\
&& \quad\quad \frac{i (-\pi)^{d/2}}{(2\pi )^d\Gamma (n_1)\Gamma (n_2)}\Gamma \Bigl(n_1+n_2-\frac{d}{2}\Bigr)
\int_0^1 d\alpha \alpha^{n_1-1}(1 - \alpha )^{n_2 - 1}
\delta \bigl(x- a \alpha - b (1-\alpha)\bigr) A^{\frac{d}{2}-n_1-n_2} \nonumber \\
\ee
and
\be
&& \int\frac{d^d k}{(2\pi)^d}\frac{\delta\left(1-\frac{n\cdot k}{n\cdot P}\right)}
{[k^2-m_1^2]^{n_1}[(k - a P)^2 - m_2^2]^{n_2}[(k - b P)^2 - m_3^2]^{n_3}} = \nonumber \\
&& \quad\quad \frac{i (-\pi)^{d/2}}{(2\pi )^d\Gamma (n_1)\Gamma (n_2)\Gamma (n_3)}
\Gamma \Bigl(n_1 + n_2 + n_3 -\frac{d}{2}\Bigr) \nonumber \\
&& \quad\quad \times\int_0^1 \{d\beta_2 d\beta_3 \}(1 - \beta_2 - \beta_3)^{n_1 - 1}
\beta_2^{n_2 - 1}\beta_3^{n_3 - 1}\delta \bigl(x - a\beta_2 - b\beta_3\bigr) B^{\frac{d}{2} - n_1 - n_2 - n_3}\,,
\ee
where the following notation is introduced:
$d=4 -2\ep$, $\{d\beta_2 d\beta_3 \} = \theta (1 - \beta_2 - \beta_3)d\beta_2 d\beta_3$ and
\be
A &=& -\alpha m_1^2 - (1-\alpha) m_2^2 \\
B &=& -\beta_2 m_2^2 - \beta_3 m_3^2 - (1-\beta_2-\beta_3) m_1^2
\ee
The masses $m_1^2$, $m_2^2$, $m_3^2$ provide infrared regularization in the cases where it is
necessary.

On the other hand these expressions for one-loop integrals could
be used as building blocks to obtain expressions for necessary
two-loop integrals.  Here we would like to note that evolution kernels, in
which we are interested, are being obtained as a result of application
$KR'$-operation to diagrams describing renormalization of effective vertices
corresponding to multiparticle distribution functions. So, in what follows we
give answers for all necessary two-loop integrals before and after application
of $KR'$-operation. Note, that it would be incorrect to use just $KR'$-subtracted
scalar integrals as extra $\ep$ terms could follow from Dirac algebra for
$\gamma$-matrices and procedure used for subtracting subdivergences provided by $R'$-operation
should take them into account.

To perform the calculation described in this paper the following general formulas
for two-loop scalar integrals are sufficient

\begin{figure*}[ht]
\vspace{-1.5cm}
\begin{center}
\begin{picture}(200,120)(0,0)
\SetScale{0.65}
\SetWidth{1}
\Line(155,109)(146,91)
\Line(154,91)(145,109)
\Line(65,-20)(140,100)
\Line(235,-20)(160,100)
\CArc(150,100)(10,0,360)
\CArc(70,107)(92.5,280,355)
\Line(210,15)(90,15)
\Vertex(212,15){4}
\Vertex(88,15){4}
\Vertex(160.5,98){3}
\end{picture}
\end{center}
\vspace{5mm}
\label{Vintegral}
\end{figure*}
\be
V(b|c,d|1) &=& -(4\pi)^4\int\frac{1}{(l - c P)^2(l - d P)^2}
\int\frac{\delta\left(x - \frac{k\cdot n}{P\cdot n}\right)}{(k - l)^2(k - b P)^2}
\frac{d^d\bar l d^d\bar k}{(2\pi)^{2d}}
\ee
Before $KR'$-operation we have
\be
V(b|c,d|1) &=& \frac{1}{2\ep y_2}\left\{
\theta (-x_1)\theta (x_1 > y_1)\widetilde{J}\left(\frac{x_1}{y_1}\right)
- \theta (-x_1)\theta (x_1 > y_1 + y_2)\widetilde{J}\left(\frac{x_1}{y_1 + y_2}\right)
\right. \nonumber \\ &&\left.
 +\  \theta (x_1)\theta (x_1 < y_1 + y_2)\widetilde{J}\left(\frac{x_1}{y_1 + y_2}\right)
- \theta (x_1)\theta (x_1 < y_1)\widetilde{J}\left(\frac{x_1}{y_1}\right)
\right\}
\ee
Here we have introduced the notation
\be
d^d\bar k &=& M(\ep )d^d k,\quad x_1 = x -b,\quad y_1 = d - b,\quad y_2 = c - d > 0 \nonumber \\
\widetilde{J}(z) &=& -\frac{1}{\ep}\ln z + \frac{1}{2}\ln^2 z,\qquad\quad
\widetilde{S}(z)\ =\  \frac{1}{\ep}\frac{1-z}{z} - 2\frac{1-z}{z} - \frac{1+z}{z}\ln z
\nonumber \\
J(z) &=& \frac{1}{\ep}\ln z + \frac{1}{2}\ln^2 z ,\qquad\qquad\!
S(z)\ =\  -\frac{1}{\ep}\frac{1-z}{z} - 2\frac{1-z}{z} - \frac{1+z}{z}\ln z \nonumber
\ee
The result for $V(b|c,d|1)$ after $KR'$-operation could be obtained from the result without
$KR'$-operation performed with the substitution $\widetilde{J}(z)\to J(z)$.
\be
V\left(b\Big|c,d\Big|\frac{l\cdot n}{P\cdot n}\right) &=&
-(4\pi)^4\int\frac{l\cdot n/P\cdot n}{(l - c P)^2(l - d P)^2}
\int\frac{\delta\left(x - \frac{k\cdot n}{P\cdot n}\right)}{(k - l)^2(k - b P)^2}
\frac{d^d\bar ld^d\bar k}{(2\pi)^{2d}}
\ee
Before $KR'$-operation we have
\be
V\left(b\Big|c,d\Big|\frac{l\cdot n}{P\cdot n}\right) &=& \frac{1}{2\ep y_2}
\left\{
\theta (x_1 > y_1)\theta (-x_1)\left[
x_1 \widetilde{S}\left(\frac{x_1}{y_1}\right) + b \widetilde{J}\left(\frac{x_1}{y_1}\right)
\right]\right. \nonumber \\ && \left.
-\ \theta (x_1 > y_1 + y_2)\theta (-x_1)\left[
x_1 \widetilde{S}\left(\frac{x_1}{y_1+y_2}\right) + b \widetilde{J}\left(\frac{x_1}{y_1 + y_2}\right)
\right]\right. \nonumber \\ && \left.
+\  \theta (x_1 < y_1 + y_2)\theta (x_1)\left[
x_1 \widetilde{S}\left(\frac{x_1}{y_1+y_2}\right) + b \widetilde{J}\left(\frac{x_1}{y_1 + y_2}\right)
\right] \right. \nonumber \\ && \left.
-\ \theta (x_1 < y_1)\theta (x_1)\left[x_1 \widetilde{S}\left(\frac{x_1}{y_1}\right)
+ b \widetilde{J}\left(\frac{x_1}{y_1}\right)\right]
\right\}
\ee
The result for $V\left(b|c,d|\frac{l\cdot n}{P\cdot n}\right)$ after $KR'$-operation could be obtained from the result without
$KR'$-operation performed with the substitutions: $\widetilde{J}(z)\to J(z)$ and
$S(z)\to \widetilde{S}(z)$
\begin{figure*}[ht]
\vspace{-1.5cm}
\begin{center}
\begin{picture}(205,120)(0,0)
\SetScale{0.65}
\SetWidth{1}
\Line(155,109)(146,91)
\Line(154,91)(145,109)
\Line(65,-20)(140,100)
\Line(235,-20)(160,100)
\CArc(150,100)(10,0,360)
\CArc(150,115)(118,238,302)
\CArc(150,-85)(118,58,122)
\Vertex(212,15){4}
\Vertex(88,15){4}
\end{picture}
\end{center}
\vspace{5mm}
\label{Wintegral}
\end{figure*}
\be
W\left(a,b\ \Big|\ c\ \Big|\left\{\begin{array}{c} 1 \\[3mm] (l\cdot n/P\cdot n)\end{array}\right\}
 \right) &=&
- (4\pi)^4\int\frac{\delta (x - \frac{k\cdot n}{P\cdot n})}{(k - a P)^2(k - b P)^2} \nonumber \\
& \times & \int\frac{1}{(l - k)^2 (l - c P)^2}\left\{
\begin{array}{c}
2 \\[3mm] (l\cdot n/P\cdot n)
\end{array}\right\}
\frac{d^d\bar k d^d\bar l}{(2\pi)^{2d}}\ .
\ee
Before $KR'$-operation we have
\be
&& \hspace*{-20mm} W\left(a,b\ \Big|\ c\ \Big|\left\{\begin{array}{c} 1 \\[3mm] (l\cdot n/P\cdot n)\end{array}\right\}
\right) \nonumber \\[3mm] &
\begin{array}{c}
a \neq b \\[2mm] =
\end{array}&\hspace*{-5mm}
\frac{1}{2\ep}\left\{\begin{array}{c} 1 \\[3mm] \frac{1}{2}(x_1+y+1) + b\end{array}\right\}
\frac{1}{y_1+y_2}\left\{\theta (x_1)\theta (y_1 + y_2 > x_1)\left[-\frac{1}{\ep} +2\right] \right. \nonumber \\
&+&\hspace*{-10mm} \theta (x_1)\theta (y_1 > x_1)\ln\frac{x_1}{y_1}
- \theta (x_1)\theta (y_1 > x_1)\theta (x_1 > y_1 + y_2)
\ln\frac{[y_1+y_2]-x_1}{[y_1+y_2]-y_1} \nonumber \\
&+&\hspace*{-10mm} \left. \theta (x_1 > y_1)\theta (y_1 + y_2 > x_1)\ln\frac{[y_1+y_2]-x_1}{[y_1+y_2]-y_1}
- \theta (-x_1)\theta (x_1 > y_1)\ln\frac{x_1}{y_1} \right\} \nonumber \\
& \begin{array}{c}
a - b \neq c \\[3mm] =
\end{array}&\hspace*{-2mm}
\frac{1}{2\ep}\left\{
\begin{array}{c}
1 \\[3mm] \frac{1}{2}(x_1 + y_1 + b)
\end{array}\right\}
\Bigl(\theta (-x_1)\theta (x_1 > y_1) + \theta (x_1)\theta (y_1 > x_1)\Bigr)
\left(
\frac{1}{y_1} + \frac{1}{x_1}
\right),
\ee
where $x_1 = x - b$, $y_1 = c - b$, $y_2 = a - c$, $y_1 + y_2 = a - b \geq 0$, $x > 0$.
After $KR'$-operation was performed the integral above takes the following form
\be
&&\hspace*{-30mm} W\left(a,b\ \Big|\ c\ \Big|\left\{\begin{array}{c} 1 \\[2mm] (l\cdot n/P\cdot n)\end{array}\right\} \right) \nonumber \\ 
&
\begin{array}{c}
a \neq b \\[2mm] =
\end{array}&\hspace*{-5mm}
\frac{1}{2\ep}\left\{\begin{array}{c} 1 \\[3mm] \frac{1}{2}(x_1+y+1) + b\end{array}\right\}
\frac{1}{y_1+y_2}\left\{\theta (x_1)\theta (y_1 + y_2 > x_1)\left[\frac{1}{\ep} +2\right] \right. \nonumber \\
& +&\hspace*{-10mm} \theta (x_1)\theta (y_1 > x_1)\ln\frac{x_1}{y_1}
- \theta (x_1)\theta (y_1 > x_1)\theta (x_1 > y_1 + y_2)
\ln\frac{[y_1+y_2]-x_1}{[y_1+y_2]-y_1} \nonumber \\
&+&\hspace*{-10mm} \left.\theta (x_1 > y_1)\theta (y_1 + y_2 > x_1)\ln\frac{[y_1+y_2]-x_1}{[y_1+y_2]-y_1}
- \theta (-x_1)\theta (x_1 > y_1)\ln\frac{x_1}{y_1} \right\} \nonumber \\
& \begin{array}{c}
a - b \neq c \\[3mm] =
\end{array}&\hspace*{-2mm}
\frac{1}{2\ep}\left\{
\begin{array}{c}
1 \\[3mm] \frac{1}{2}(x_1 + y_1 + b)
\end{array}\right\}
\Bigl(\theta (-x_1)\theta (x_1 > y_1) + \theta (x_1)\theta (y_1 > x_1)\Bigr)
\left(
\frac{1}{y_1} + \frac{1}{x_1}
\right).
\ee

\section*{Appendix C}

Here, we following Refs.~\cite{MPhi3,MQEDNS}, present the details for the derivation of non-diagonal
part of anomalous dimensions matrix for fermionic operator, based on the analysis of conformal
Ward identity for Green functions with fermionic operator insertion. In what follows, we will
consider somewhat more general fermionic operator, then one considered in the main
body of the paper:
\be
O_{j,l}^{\mathbf {fer}} = \bar\psi_{+}(i\partial_{+})^{l}P_{j}^{(\alpha,\beta )}
\left(
\frac{D_{+}}{\partial_{+}}
\right)\phi\, .
\ee
Even so, this operator will be highest vector in corresponding conformal
representation only for $\alpha =1$ and $\beta = 0$, some of the formulas
given here will certainly have applications in other supersymmetric models
with more rich field content. To start with, let us remained reader some
basic facts about conformal symmetry group.

Conformal group is defined as the most general group leaving invariant
light-cone in $n$-dimensional Minkowski space. The algebra of this group
contains, in addition to generators of Poincare group: $P_{\mu}$ and $M_{\mu\nu}$,
the generator of dilations $D$ and $n$ generators $K_{\mu}$ of special
conformal transformations. The latter generators satisfy the following commutation
relations~\cite{Mack}:
\be
[D,K_{\mu}] &=& iK_{\mu}\, , \qquad \;\;\;\;\,
[K_{\mu},P_{\nu}]\  =\  -2i(g_{\mu\nu}D+M_{\mu\nu}), \nonumber \\[1mm]
 [D,P_{\mu}] &=& -iP_{\mu}\, , \qquad 
[K_{\rho},M_{\mu\nu}]\  =\  i(g_{\rho\mu}K_{\nu}-g_{\rho\nu}K_{\mu}), \nonumber \\[1mm] 
 [K_{\mu},K_{\nu}] &=& 0\, , \qquad \qquad\, [D,M_{\mu\nu}]\  =\  0\, .
\ee
Here we will work only with the subalgebra of conformal algebra, the so called
collinear conformal algebra, isomorphic to $SU(1,1)\cong SO(2,1)$. The representations
of collinear conformal algebra could be obtained from those of full conformal algebra
via a projection to the light cone. The necessary projection itself could be build with
the help of two light-cone vectors, $n$ and $n^{*}$ with the properties:
$n^2 = n^{*2} = 0$ and $n\cdot n^* = 1$.  The generators of collinear conformal
group obtained have the following expressions:
\be
K_{-} &=& n^{*}_{\mu}K^{\mu},\ D\,, \nonumber \\
P_{+} &=& n_{\mu}P^{\mu},\ M_{- +}\  =\  n^{*}_{\mu}M^{\mu\nu}n_{\nu}\,.
\ee
In what follows, we will also need the transformation laws of fermion and scalar fields
under special conformal transformations:
\be
K_{-}\phi &=& i\bigl(2x_{-}(d_{\phi}+x\cdot\partial)-x^2\partial_{-}\bigr)\phi, \nonumber \\
K_{-}\psi &=& i\bigl(2x_{-}(d_{\psi}+x\cdot\partial)
+2\Sigma_{-\nu}x^{\nu}-x^2\partial_{-}\bigr)\psi
\ee
where $d_{\phi}$ and $d_{\psi}$ are scale dimensions for the scalar and fermion fields
correspondingly. $\Sigma_{-\nu}$ is the spin operator:
\be
\Sigma_{\mu\nu}\psi = \frac{1}{4}[\gamma_{\mu},\gamma_{\nu}]\psi.
\ee
The fermion conformal operators, we are interested in, form an infinite dimensional
representation of collinear group. The members of each series are labeled by their
spin $l$. The generators $D$ and $M_{- +}$ are diagonal with respect to representations,
whereas $P_{+}$ acts as a raising operator and $K_{-}$ as a lowering operator.
The action of $K_{-}$ on $O_{j,l}^{\mathbf {fer}}$ is
\be
K_{-}O_{j,l}^{\mathbf {fer}} = \int d^4 x \left\{
\left(K_{-}\phi (x)\right)\frac{\delta}{\delta\phi (x)}
+ \left(K_{-}\bar\psi (x)\right)\frac{\delta}{\delta\bar\psi (x)}
\right\}O_{j,l}^{\mathbf {fer}}.
\ee
Now, it is an easy exercise to find that
\be
K_{-}O_{j,l}^{\mathbf {fer}} = a(j,l,\alpha,\beta)O_{j,l-1}^{\mathbf {fer}},
\ee
where
\be
a(j,l,\alpha,\beta) = 2(j-l)(j+l+\alpha +\beta +1).
\ee
To derive the conformal Ward identity itself we start with generating functional
for renormalized disconnected Green functions with operator insertions
\be
Z_{j,l}(\bar\eta,\eta,J) &=& \frac{1}{N}\int D\Phi [O_{j,l}^{\mathbf {fer}}]\cdot
\exp\left\{
i[S] + i\int d^d x [\bar\eta (x)\psi (x) + \bar\psi (x)\eta (x) + J(x)\phi (x)]
\right\}, \nonumber \\
 i[S] &=& i\int d^d x {\cal L}\,,
\ee
where $[O_{j,l}^{\mathbf {fer}}]$ and $[S]$ denote corresponding renormalized
quantities. Using the invariance of the generating functional with respect to
special conformal transformations and performing differentiation over field
sources, one gets the following conformal Ward identity
\be
\langle [O_{j,l}^{\mathbf {fer}}] \left(\delta^K\chi \right)\rangle =
- \langle\left(\delta^K [O_{j,l}^{\mathbf {fer}}]\right)\chi\rangle
-\langle [O_{j,l}^{\mathbf {fer}}]\left(\delta^K [S]\right)\chi\rangle,
\label{CWI}
\ee
where $\langle A\rangle$ stands for vacuum averaging of the time ordered product
$TA\exp (i[S])$ and $\chi = \Pi_i\phi_i$ denotes product of elementary fields
taken at different space-time points. Note, that the left-hand side of Eq.~(\ref{CWI}) is finite, so the
right-hand side also should be finite. The first term at right-hand side is a variation
of the renormalized fermion operator under special conformal transformations. It is
given by
\be
\delta^K [O_{j,l}^{\mathbf {fer}}] = i\sum_{k=0}^j
\left\{
\hat Z\hat A(l,\ep)\hat Z^{-1}
\right\}_{j,k}[O_{k,l-1}^{\mathbf {fer}}], \label{deltaop}
\ee
where we introduced the following notation
\footnote{ The renormalization of conformal operators does not depend on $l$ due
to Lorentz invariance and the matrix $Z_{jk}$ is triangular.}
\be
[O_{j,l}^{\mathbf {fer}}] = \sum_{k=0}^{j}Z_{jk}(\ep,g)O_{k,l}^{\mathbf {fer}},
\qquad k\leq l\,,
\ee
and\footnote{Note, that here we used the fact that anomalous dimensions of scalar
and fermion fields are equal, as they  belong to the same supermultiplet.}
\be
\hat A(l,\ep) &=& \hat a(l) + 2 (\gamma_{\Phi}+\ep)\hat b(l), \\
\hat a(l) &=& \bigl\{a(j,l,\alpha,\beta)\delta_{jk}\bigr\}, \\
\hat b(l) &=& 2\int_{-1}^{1} dx \frac{w(x|\alpha,\beta)}{{\cal N}n_k(\alpha,\beta)}
P_k^{(\alpha,\beta)}(x)\hat{\cal L}P_j^{(\alpha,\beta)}(x),
\ee
with
\be
\hat{\cal L} &=& l - x\frac{d}{dx},\quad w(x|\alpha,\beta) = (1-x)^{\alpha}(1+x)^{\beta}, \nonumber \\
{\cal N}_k(\alpha,\beta) &=& 2^{\alpha +\beta +1}
\frac{\Gamma (k+\alpha +1)\Gamma (k+\beta +1)}{(2k+\alpha+\beta+1)\Gamma (k+1)\Gamma (k+\alpha +\beta +1)}\,.
\ee
The evaluation of integral in the definition of matrix $\hat b(l)$ gives
\be
b_{jk}(l) &=& 2 (l-k)\delta_{jk} \nonumber \\
&-& \theta (j>k)\left\{
(2k+\alpha+\beta + 1)\frac{\Gamma (k+\alpha +\beta +1)}{\Gamma (j+\alpha +\beta +1)}
\left[
(-1)^{j-k}\frac{\Gamma (j+\alpha+1)}{\Gamma (k+\alpha +1)}
+\frac{\Gamma (j+\beta +1)}{\Gamma (k+\beta +1)}
\right]
\right\}. \nonumber
\ee
The second term on the right-hand side of Eq.~(\ref{CWI}) is the product of two renormalized
operators: $[O_{j,l}^{\mathbf {fer}}]$ and the variation of renormalized action. For the latter
we have the following expression
\be
i\delta_{-}^K[S] = \frac{\bar\beta (\ep,g)}{g}[\triangle_{-}]\quad \mbox{with}\quad
[\triangle_{-}] = g\frac{\partial}{\partial g}i\int d^d x 2x_{-}{\cal L}(x), \quad
\bar\beta (\ep, g) = -\ep g + \beta (g). \label{deltaS}
\ee
As it is known, the product of two renormalized operators may still contain ultraviolet
divergences, so an additional subtraction of divergences is necessary
\be
[O_{j,l}^{\mathbf {fer}}\triangle_{-}] = [O_{j,l}^{\mathbf {fer}}][\triangle_{-}]
-i\sum_{k=0}^{j}Z_{jk}^{*}(l)[O_{k,l-1}^{\mathbf {fer}}]. \label{productop}
\ee
Substituting Eqs.~(\ref{deltaop}), (\ref{deltaS}) and (\ref{productop}) into conformal
Ward identity and taking limit $\ep\to 0$, one gets\footnote{Note, that at this point
the finiteness of the right-hand side of CWI is used }
\be
\langle [O_{j,l}^{\mathbf {fer}}]\delta^K_{-}\chi\rangle =
i\sum_{k=0}^{j}\{\hat a(l) +\hat\gamma^c (g,l) \}_{jk}\langle [O_{k,l}^{\mathbf {fer}}]\chi\rangle
- \frac{\beta}{g}\langle [O_{j,l}^{\mathbf {fer}}\triangle_{-}] \rangle.
\ee
Here the definition for the matrix of special conformal anomalies was introduced
\be
\hat\gamma^c = \lim_{\ep\to 0}\left(\hat Z\hat A\hat Z^{-1} -\hat a -\frac{\bar\beta}{g}
\hat Z^{*} \right).
\ee
It is easy to derive, that in Wess-Zumino model we are considering here $\hat Z^{*[1]}(l)$
(the coefficient of simple pole in $\ep$) is given by
\be
\hat Z^{*[1]}(l) = g^2\hat\gamma^{(1)}\hat b(l) + {\cal O}(g^2)
\ee
and as a consequence we have $\hat\gamma^c = \hat b(l)(\hat\gamma^{(1)}+2\gamma_{\Phi}^{(1)}\hat 1)$
with $\hat 1 = \{\delta_{jk}\}$.

Now, the crucial point is to note, that the scale and special conformal anomalies are in fact related.
The relation between matrices of scale and special conformal anomalies could be obtained by
considering the action of the commutator of special conformal transformation with dilation
$[\delta^D,\delta^K_{-}] = \delta^K_{-}$ on Green function with fermion operator insertion.
Taking into account that $\delta^D$ acts on $\langle [O_{j,l}^{\mathbf {fer}}]\chi\rangle$
as $\mu\frac{\partial}{\partial\mu}+N_{\psi}\gamma_{\psi}+N_{\phi}\gamma_{\phi}$
($N_{\psi}$ and $N_{\phi}$ are numbers of $\psi$ and $\phi$ fields in $\chi$) one gets the
following identity
\be
\left[\hat a(l)+\hat\gamma^c (g,l) + 2\frac{\beta (g)}{g}\hat b(l), \hat\gamma (g)\right] = 0\,.
\ee
Since $\hat a(l)$ does not depend on coupling constant, solving this equation recursively
we see that $\gamma^{ND}_{jk}$  in $(n)$-loop order is determined by the $(n-1)$-loop
approximation of $\hat\gamma^c$, $\beta$ and $\hat\gamma = \hat\gamma^{D}+\hat\gamma^{ND}$.
At two first order we have
\be
\gamma^{(0)}_{jk} &=& \delta_{jk}\gamma^D_k, \\[1mm]
[\gamma^{ND}_{jk}]^{(1)} &=& \frac{\gamma^{(0)}_j -\gamma^{(0)}_k}{2(j-k)(j+k+2)}
(\gamma^{c(0)}_{jk}-\beta_0b_{jk})\,.
\ee
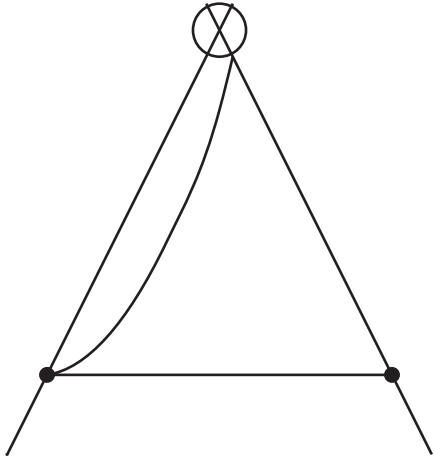
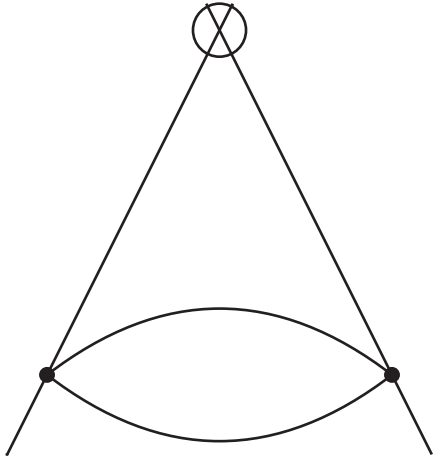
\end{document}